\documentclass[11pt,a4paper]{article}

\usepackage{amsmath,amssymb}
\usepackage{epsfig,graphicx}
\usepackage{subfigure}
\usepackage{graphicx}
\usepackage{rotating}
\usepackage{cancel}
\usepackage{bm}
\usepackage{color}
\usepackage{comment}
\usepackage[nosort]{cite}
\usepackage{psfrag}
\usepackage{overpic}
\usepackage{bbm}

\newcommand{\varix}{{\varkappa}}

\renewcommand\[{\left[}

\newcommand{\exclude}[1]{}

\def\beq{\begin{equation}}
\def\eeq{\end{equation}}

\begin{document}
\numberwithin{equation}{section}
\title{
\vspace{2.5cm} 
\Large{\textbf{Monodromy Dark Matter
\vspace{0.5cm}}}}

\author{Joerg Jaeckel, Viraf M. Mehta and Lukas T.~Witkowski\\[2ex]
\small{\em Institut f\"ur Theoretische Physik, Universit\"at Heidelberg,} \\
\small{\em Philosophenweg 16, 69120 Heidelberg, Germany}\\[0.5ex] }

\date{}
\maketitle

\begin{abstract}
\noindent
Light pseudo-Nambu-Goldstone bosons (pNGBs) such as, e.g.~axion-like particles, that are non-thermally produced via the misalignment mechanism are promising dark matter candidates. An important feature of pNGBs is their periodic potential, whose scale of periodicity controls all their couplings. As a consequence of the periodicity the maximal potential energy is limited and, hence, producing the observed dark matter density poses significant constraints on the allowed masses and couplings.
In the presence of a monodromy, the field range as well as the range of the potential can be significantly extended.
As we argue in this paper this has important phenomenological consequences. The constraints on the masses and couplings are ameliorated and couplings to Standard Model particles could be significantly stronger, thereby opening up considerable experimental opportunities.
Yet, monodromy models can also give rise to new and qualitatively different features. As a remnant of the periodicity the potential can feature pronounced ``wiggles''. When the field is passing through them quantum fluctuations are enhanced and particles with non-vanishing momentum are produced. Here, we perform a first analysis of this effect and delineate under which circumstances this becomes important. We discuss possible cosmological consequences.
\end{abstract}

\newpage

\section{Introduction}
Although more than three quarters of a century have passed since Zwicky~\cite{Zwicky:1933gu} found the first evidence for Dark Matter (DM),
its nature and even basic properties such as the mass of DM objects are essentially unknown.
Possible masses range from ultralight particles $10^{-22}\,{\rm eV} \lesssim m_{\rm DM}$ (cf.~\cite{Marsh:2015xka}) to seriously macroscopic objects with $10^{-7}M_{{\rm solar}}\sim 10^{59}\,{\rm eV}$ or even more~(see, e.g., \cite{Carr:2009jm,Capela:2013yf}) which clearly shows the level of our ignorance. 

While experimental searches for a number of candidate particles, such as WIMPs (cf.~\cite{Bertone:2004pz,Agashe:2014kda} for reviews), axions (cf.~\cite{Sikivie:2006ni,Marsh:2015xka} for reviews) are ongoing and even entering the most promising parameter regions, these searches rely on specific properties of the putative DM particles. DM particles with different properties could be missed. For example axion dark matter experiments are essentially insensitive to WIMPs and vice versa. 
In order not to exclude anything due to theoretical bias it is therefore prudent to also consider new candidates as well as to re-examine the properties of existing ones in the light of new theoretical developments. This is the spirit we will adopt in the present paper.

A large class of interesting DM particles are pseudo-Nambu-Goldstone bosons of spontaneously broken approximate global symmetries, with the axion~\cite{Peccei:1977hh, Peccei:1977ur, Wilczek:1977pj, Weinberg:1977ma,Preskill:1982cy,Abbott:1982af,Dine:1982ah} being perhaps the most prominent example. Other well motivated examples include
axion-like particles~\cite{Masso:1995tw,Masso:1997ru,Masso:2004cv,Jaeckel:2006xm,Arias:2012az}, familons~\cite{Wilczek:1982rv,Joshipura:1987tf,Berezhiani:1989fp,Carone:2012dg,Jaeckel:2013uva}, and there are many more.
In section~\ref{monodromy} we will briefly recall the essential features of DM consisting of pNGBs and produced via the misalignment mechanism. In particular, we will review that the periodicity puts an upper limit on the amount of DM of this type. In addition, note that a single pNG species can be DM only for certain combinations of the periodicity
and mass. As couplings to other particles including those of the Standard Model are linked to the periodicity, this puts significant restrictions in the space of couplings and masses. We then show that if the DM field exhibits a (axion) monodromy \cite{0803.3085, 0808.0706, 0811.1989},
these restrictions can be lifted and a wide range of new parameter space opens up (cf.~Fig.~\ref{parameterspace}).\footnote{While writing this paper, the possibility of a monodromy in a theory of dark energy interacting with dark matter has appeared in \cite{1605.00996}.} A similar increase in parameter space was investigated in \cite{Higaki:2016yqk} where the authors considered an aligned QCD-like axion in order to enhance the allowed region.

This enhancement is phenomenologically important as the parameter regions in question are at larger coupling and therefore accessible to a wide range of near future experiments and techniques~\cite{Asztalos:2009yp,Heilman:2010zz,Rybka:2014xca,Nguyen:2015ktw,Semertzidis2016,Graham:2011qk,Horns:2012jf,Jaeckel:2013eha,Suzuki:2015sza,Suzuki:2015vka,Veberic:2015yua,Redondo2016,Sikivie:2013laa,Graham:2013gfa,Budker:2013hfa,Rybka:2014cya,VanTilburg:2015oza,Graham:2015ifn}. 

While the field space is enlarged by the monodromy, the original periodicity may still have phenomenological consequences. The essential remnant of this
is a periodic contribution to the potential. As a result, the potential has ``wiggles'' that can vary in size depending on the choice of parameters. Sufficiently small wiggles leave the evolution more or less unaffected compared to a simple quadratic potential. However, the bigger the periodic part of the potential the larger the deviations in the evolution. On a first level the classical evolution of the homogeneous field value will be modified. We will investigate this in section~\ref{classical}. In the case of pronounced wiggles we obviously have regions where the curvature of the potential is negative. Quantum-mechanically, such instabilities are linked to the growth of fluctuations and, in effect, particle production.\footnote{The effect of such modulations on the primordial power spectrum, in the context of large-field inflation, has been studied in \cite{Choi:2015aem}.} In section~\ref{quantum} we will make a first investigation of this effect.
The classical and, in particular, the quantum evolution exhibit a very rich behaviour and our investigation here 
can only be viewed as a first step. In the final section~\ref{conclusions} we therefore outline directions for further investigations.

\section{From pseudo-Nambu-Goldstone to Axion Monodromy Dark Matter}\label{monodromy}

The phenomenology of pseudo-Nambu-Goldstone bosons is strongly affected by the following two observations. For one, their mass can be quite small due to protection by symmetry. Second, their couplings are suppressed by the scale of spontaneous symmetry breaking:
\begin{itemize}
\item{} The mass is given by $m^2\sim \frac{\Lambda^4}{f}$, where $f$ is the scale of spontaneous symmetry breaking and $\Lambda$ quantifies a (small) explicit breaking.
\item{} Couplings take the form $\sim \frac{\phi}{f} F\tilde{F}$, $\sim \frac{\partial_{\mu}\phi}{f}\bar{\psi}^{\prime}\gamma^{\mu}\gamma^{5}\psi$, etc.~and are suppressed by $f$.
\end{itemize}

\subsection{Pseudo-Nambu-Goldstone Dark Matter}
In the context of cosmology, light scalar and pseudo-scalar fields provide excellent candidates for Dark Matter when produced by the misalignment mechanism (see e.g.~\cite{Arias:2012az} for details).
The simplest realisation of this arises when the light field is already present during inflation. In this case the field will have a homogeneous value all across the observable Universe. However, this field value is not necessarily zero as
the field may not have had enough time to relax to zero. Indeed, this is the generic case as Hubble friction in the equation of motion will prevent significant rolling of the field as long as $H\gg m$:
\begin{equation}
\label{motion}
\ddot{\phi}+3H\dot{\phi}+m^{2}\phi=0,
\end{equation}
It is straightforward to check that once and only once $H\ll m$ the field starts oscillating and its energy density behaves like,
\begin{equation}
\label{density}
\rho_{\phi}(t)\sim \left(\frac{a_{1}}{a(t)}\right)^3\rho_{\phi,1}\sim \left(\frac{a_{1}}{a(t)}\right)^3 m^2\phi^2_{1},
\end{equation}
where the quantities with index $1$ are evaluated roughly when $H\sim m$. This is exactly the dilution by a volume factor that one expects for matter.

\begin{figure} 
\centering
\includegraphics[width=0.60\textwidth]{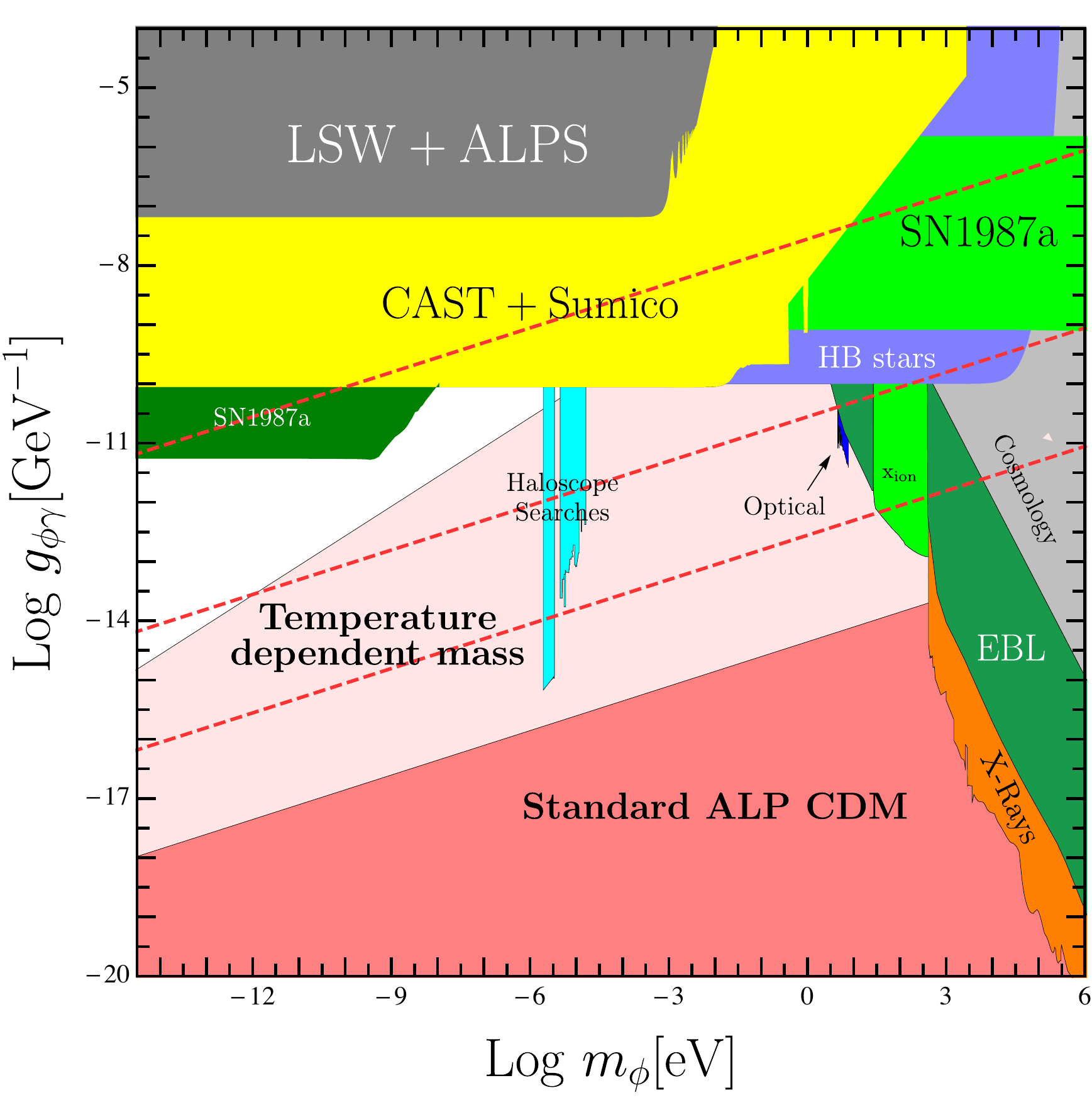}
  \caption{Plot of mass versus couplings for axion-like particles coupled to two photons. The regions where standard pNGBs with constant mass terms provide for enough dark matter are indicated by the red shaded area.
Regions which may be reached with a suitable temperature dependence of the mass are indicated by transparent red.
Monodromy Dark Matter can provide for sufficient density in the extended regions indicated by the dashed lines allowing (from bottom to top) for initial field values up to $\phi_{\mathrm{initial}}= 2 \pi \cdot (10,\,10^{3},\,10^{6})f$.  
Plots are adapted from~\cite{Arias:2012az,Dobrich:2015jyk,Jaeckel:2015jla} where also the experimental and observational constraints are discussed.}	
\label{parameterspace}
\end{figure}

Eq.~\eqref{density} clearly shows that the total energy density in DM is determined by the initial field value or more precisely the initial energy density.
This is where the Goldstone nature imposes restrictions. Field values of pNGBs are restricted to
\begin{equation}
\phi\lesssim 2\pi f.
\end{equation}
where $f$ is the scale of spontaneous symmetry breaking that also enters all the coupling strengths.
Pseudo-Nambu-Goldstone bosons feature a periodic potential that, by its very nature, is bounded from above,
\begin{equation}
\label{cospotential}
V(\phi)=\Lambda^4 \left[1-\cos\left(\frac{\phi}{f}\right)\right], \qquad m^{2}=\frac{\Lambda^4}{f^2}.
\end{equation}
For a constant field value this therefore limits the initial energy density to be $\rho_{\phi,1}\leq\Lambda^4$ which in turn
constrains the possible amount of DM for any given mass. The requirement of reproducing the observed amount of DM then leads to constraints on the parameter space for the pNGBs. For example, the regions shaded in red in Fig.~\ref{parameterspace} correspond to the allowed parameter space for axion-like particles coupling to two photons.

While the simple arguments above were made for a purely quadratic potential, the $\cos$-potential is approximately quadratic for the relatively small field values where most of the cosmological evolution happens. All in all, orders of magnitude changes from the limits outlined above are difficult to achieve (and would in any case require significant tuning that may even be impossible~\cite{Wantz:2009it}\footnote{If the spontaneous symmetry breaking that gives rise to the pNGB appears only after inflation, additional contributions may arise from topological defects such as strings. See~\cite{Fleury:2015aca} (and references therein) for a recent discussion.}).

One way out of this is a time-dependent or temperature-dependent potential. This is indeed the case for the QCD axion.
If the mass/potential grows with decreasing temperature the onset of the oscillations may be somewhat delayed and thereby today's energy density increased. Alternatively, from a different point of view, the mass/potential may be larger today than at the point in time when the field started oscillating, thereby giving an increase in energy density.
However, this approach has its limits (see~\cite{Arias:2012az} for details). Under the assumption that the mass of Dark Matter particles should be essentially constant from the time of matter radiation equality, the attainable region with a sufficiently high dark matter density is shown in light red in Fig.~\ref{parameterspace}. 

\subsection{Monodromy Dark Matter}
In presence of a monodromy~\cite{0803.3085, 0808.0706, 0811.1989} the degeneracy between the different minima
of the $\cos$-potential Eq.~\eqref{cospotential} is lifted.
\begin{equation}
\label{monopot}
V(\phi)=\frac{1}{2}m^2_{\mathrm{mono}}\phi^2+\Lambda^2\left[1-\cos\left(\frac{\phi}{f}+\alpha\right)\right].
\end{equation}
Here $m_{\mathrm{mono}}$ denotes the mass term arising in the monodromy that breaks the shift symmetry
\begin{equation}
\phi\to\phi+2\pi f,
\end{equation}
and $\alpha$ is an a priori arbitrary phase. The phase $\alpha$ will not have much effect on the phenomena we will study and hence we will set $\alpha=0$ in the following.

The idea of an axion with a monodromy has been exploited extensively for constructing models of large-field inflation. For the original work in string theory see \cite{0803.3085, 0808.0706} and \cite{0811.1989} in field theory.\footnote{For a review on subsequent developments until 2014 see \cite{Baumann:2014nda}. More recent advances and further references can be found in \cite{1409.5350}.} Furthermore, potentials involving an axion-like field with a monodromy have also been suggested to address aspects of the electroweak hierarchy problem~\cite{1504.07551} in a dynamical setting.\footnote{See~\cite{Jaeckel:2015txa} for a discussion of this mechanism with regard to the hierarchy problem.}

Introducing a monodromy may look like a small change in the potential, yet, it is a profound change in the symmetry structure (see also~\cite{1509.00047}). A pNGB of a spontaneously broken U(1) symmetry has an intrinsically compact field range. The shift symmetry is not global but gauged. The minima of the $\cos$-potential all correspond to exactly the same physical state. In essence, there is only one physical minimum. If this is the case, a mass term for the axion as written down in \eqref{monopot} would simply be inconsistent. 

In order to work with a consistent theory giving rise to potential Eq.~\eqref{monopot} the following conditions have to be satisfied. For one, a potential of the form \eqref{monopot} would be permissible if the shift-symmetry was a global symmetry that is then broken explicitly by the mass term.\footnote{While the existence of a UV completion is less pressing than in models of axion inflation, note that the presence of global symmetries is conjectured to be inconsistent with the existence of quantum gravity \cite{1011.5120}.}

Alternatively, a potential of the form \eqref{monopot} can arise for an axion with a multi-branched potential \cite{hep-th/0507215, 0811.1989}. Such axion theories permit the existence of an axion mass term while still adhering to an underlying shift symmetry. This can be realised by coupling the axion $\phi$ to a 3-form field $C_3$ through its field strength $F_4$ \cite{hep-th/0507215, 0811.1989} (for recent applications of this mechanism in the context of axion inflation and cosmological relaxation see \cite{1404.3040, 1512.00025, 1512.03768}):
\begin{equation}
\mathcal{L}= \tfrac{1}{2} \partial_{\mu} \phi \partial^{\mu} \phi - \tfrac{1}{4} |F_4|^4 + g \phi F_4 \ . 
\end{equation}
A 3-form field is not dynamical in 4 dimensions. Instead, the field strength $F_4$ is quantized and its values correspond to a discrete set of cosmological constants. Most importantly, one can integrate out $F_4$ through its equation of motion $\star F_4 = f_0 +g \phi$ to arrive at a theory with a potential for the axion
\begin{equation}
\label{eq:KSpot}
V= \tfrac{1}{2} (f_0 + g \phi)^2 \ .
\end{equation} 
The shift symmetry is still manifestly realized as $f_0$ is also affected by the shift:
\begin{equation}
\phi \rightarrow \phi + 2 \pi f \ , \qquad f_0 \rightarrow f_0 - 2 \pi f g \ .
\end{equation} 
This implies that there is not just one potential of type \eqref{eq:KSpot}, but rather a whole family of potentials which differ in the value for $f_0$. Models of this type have the advantage that the shift symmetry still protects the potential from potentially dangerous corrections: in particular, corrections to the potential $V$ appear as powers of $V/\Lambda^4$ with $\Lambda$ the UV cutoff \cite{1101.0026, 1507.06793}. 

Keeping the above picture in mind we now set the model building aspects aside.\footnote{In models of axion monodromy the field range for a single axion can be enhanced compared to its `decay constant' $f$ which controls its couplings to gauge bosons. A similar enhancement of the field range can also arise through a mechanism of `alignment' in a theory of more than one axion \cite{Kim:2004rp} (see also \cite{1404.7496, 1503.01015}), leading to an effective axion potential with a long period with long short period modulations. Our analysis in this paper is also applicable in this case.} Instead, we will focus on the phenomenological consequences of a field with a potential \eqref{monopot}. In particular, we will study the suitability of such fields as DM candidates. 
\bigskip

Compared to pNGBs discussed above, fields with a monodromy as in Eq.~\eqref{monopot} exhibit one crucial difference: The field values and, more importantly, the values of the potential itself are not
bounded from above.\footnote{Note that the initial field value for the axion may not be unlimited as previously discussed in the context of large field inflation and EW relaxation \cite{1512.00025}.} The simplest situation arises when the $\cos$-part of the potential is so small that it can be safely neglected. In this case the field behaves like a simple non-interacting scalar field with equation of motion as in Eq.~\eqref{motion}, but with $m$ replaced by $m_{\mathrm{mono}}$. Misalignment production of Dark Matter then proceeds along the same lines as for pNGBs, but with one crucial difference: The initial field value can now be much larger than $f$ and, therefore, 
we can obtain an essentially unbounded amount of DM. An immediate result is that today's DM density can also be achieved in the regions below the dashed lines in Fig.~\ref{parameterspace}.This significantly enlarges the permissible parameter space.

 \begin{figure}
    \centering
      \subfigure[
    ]
    {  \begin{overpic}[width=0.44\textwidth]{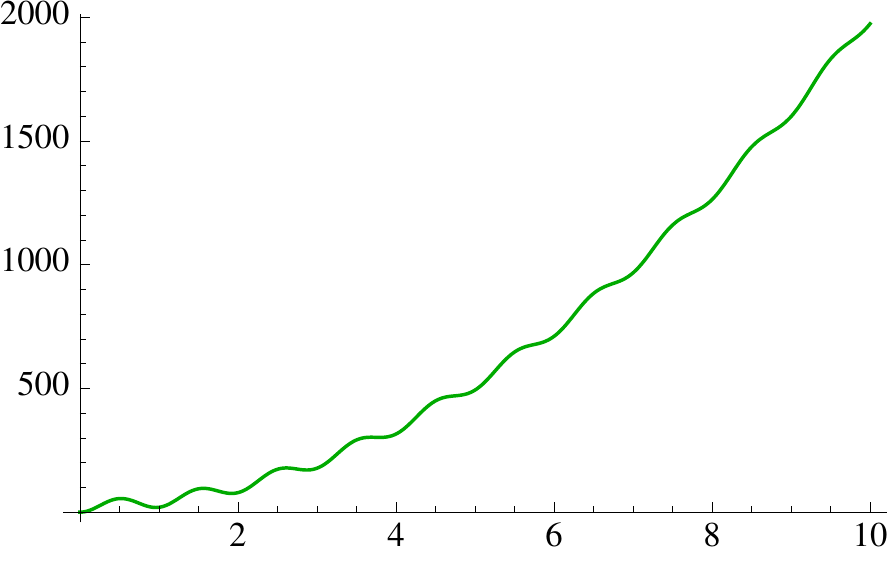}
 \put (16,62) {$\tfrac{V}{f^2 m_{\mathrm{mono}}^2}$} \put (88,-4) {$\tfrac{\phi}{2 \pi f}$} 
\end{overpic}
 }
\subfigure[
      ]
    {   \begin{overpic}[width=0.44\textwidth]{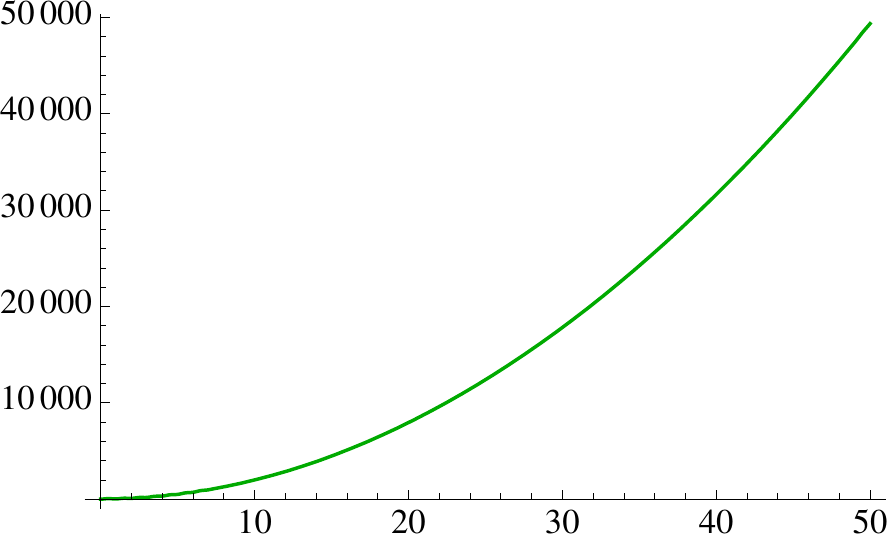}
 \put (16,62) {$\tfrac{V}{f^2 m_{\mathrm{mono}}^2}$} \put (88,-4) {$\tfrac{\phi}{2 \pi f}$} 
\end{overpic}
 }
\\
\subfigure[
      ]
 {   \begin{overpic}[width=0.44\textwidth]{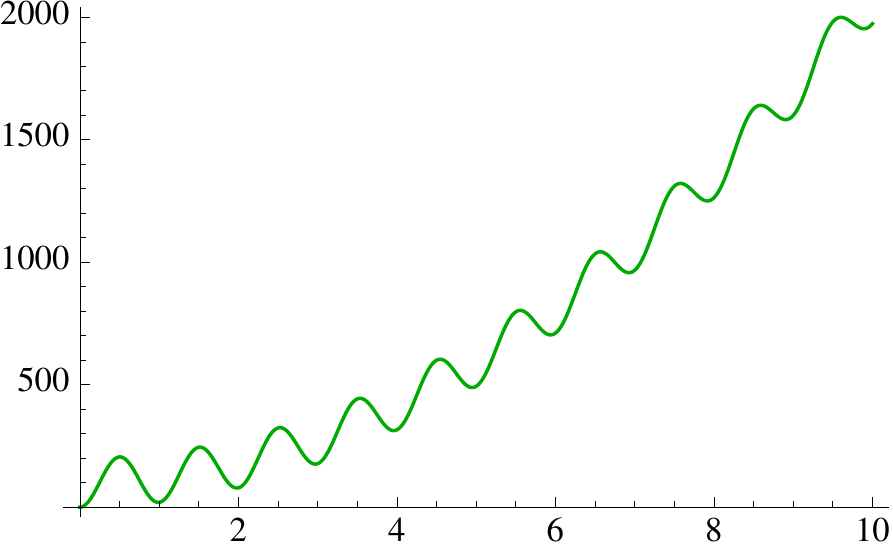}
 \put (16,62) {$\tfrac{V}{f^2 m_{\mathrm{mono}}^2}$} \put (88,-4) {$\tfrac{\phi}{2 \pi f}$} 
\end{overpic}
 }
\subfigure[
      ]
    {   \begin{overpic}[width=0.44\textwidth]{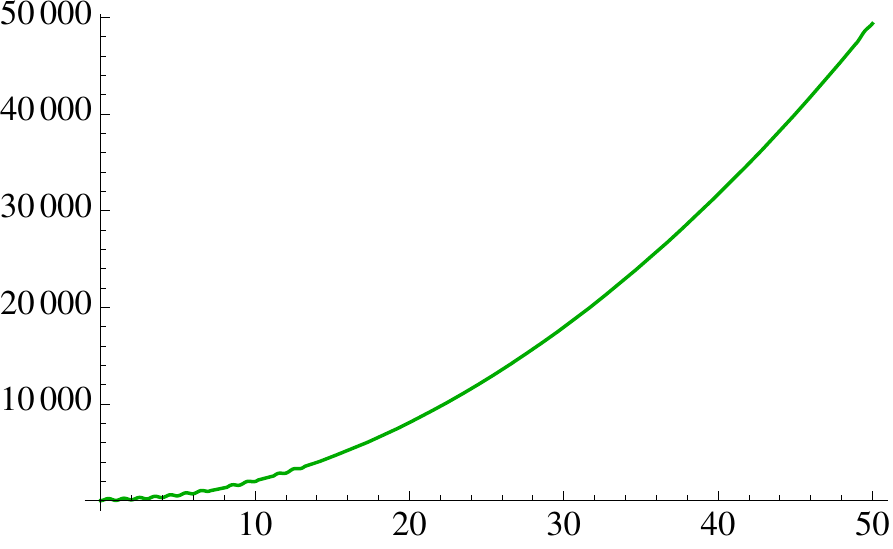}
 \put (16,62) {$\tfrac{V}{f^2 m_{\mathrm{mono}}^2}$} \put (88,-4) {$\tfrac{\phi}{2 \pi f}$} 
\end{overpic}
 }
 \centering\caption{Potential $V$ vs.~$\varphi$ for (a,b) $\kappa=5.0$ and (c,d) $\kappa=10$. Left panels show the region around the minimum where one has pronounced wiggles. The right panel shows the essentially quadratic behaviour for larger field values.}
 \label{potsplot}
  \end{figure}

\begin{figure}[t]
 \subfigure[]{
\begin{overpic}[width=0.44\textwidth]{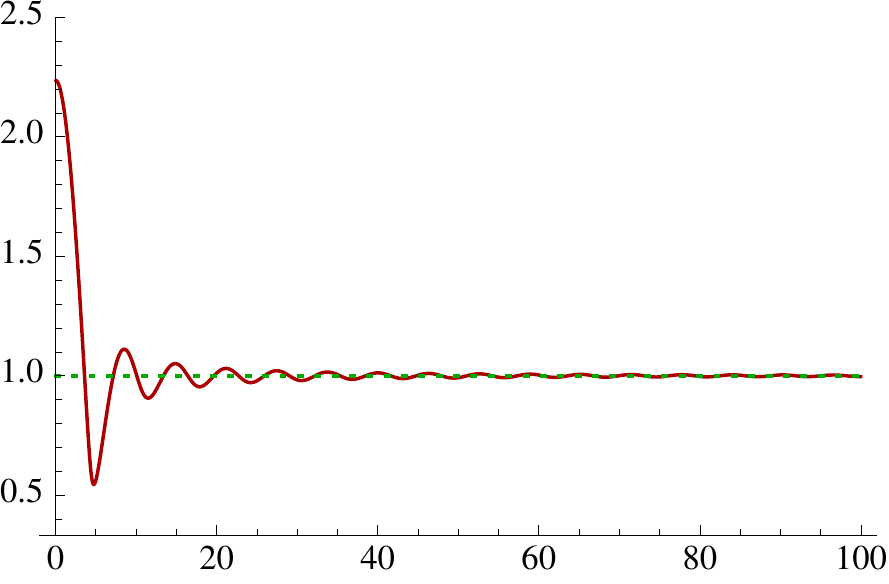}
 \put (12,62) {$\Omega$} \put (84,-4) {$\varphi_{\mathrm{initial}}$} 
\end{overpic}
 }
\subfigure[]{
 \begin{overpic}[width=0.44\textwidth]{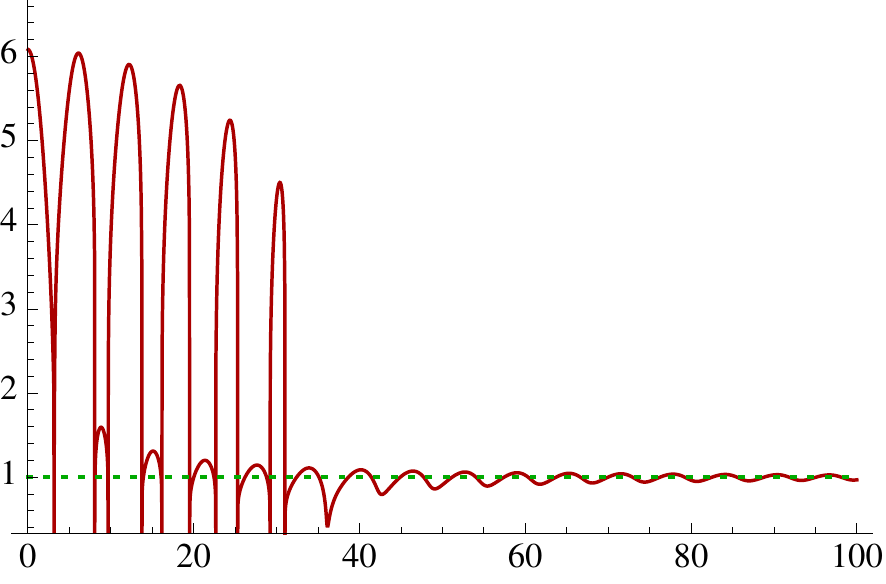}
 \put (10,62) {$\Omega$} \put (84,-4) {$\varphi_{\mathrm{initial}}$} 
\end{overpic}
 }
\caption{Angular oscillation frequency $\Omega$ for (a) $\kappa=2.0$ and (b) $\kappa=6.0$ as a function of $\varphi_{\mathrm{initial}}$. Note that for $\kappa=6.0$ the potential has additional local minima while there is only one global minimum for $\kappa=2.0$. The peaks in the angular frequency for the case $\kappa=6.0$ arise when the field oscillates about one of the local minima.}
\label{fig:frequency}
\end{figure}

While technically nearly trivial, this is still a central result of this paper. Light DM particles could exist in regions of parameter space where the scale $f$ is smaller and the coupling to matter is stronger. Thereby even experiments that do not yet have sensitivity to the simplest pNG DM models have discovery potential for monodromic DM.

To conclude this section let us note that we have assumed higher order shift symmetry breaking terms, e.g.~$\sim \phi^4$ to be absent. At large field values such terms would modify the evolution of the scalar field and change the equation of state away from that of dark matter. In some cases (e.g.~a a pure $\phi^4$ term) this again limits the amount of dark matter available today. Whether higher terms in the field $\phi$ are present will depend on the realization of this Dark Matter sector in a more complete model. For example, note that if we take $\phi$ to be an axion with a multibranched potential from a coupling to a 3-form field, one can ensure that higher shift symmetry breaking terms are only radiatively generated and severely suppressed.

\section{Classical evolution}\label{classical}
So far we have assumed that the $\cos$-term in the monodromy potential~\eqref{monopot} can be neglected.
However this is not necessarily the case.
The importance of the $\cos$-term can be quantified by the parameter,
\begin{equation}
\kappa^2=\frac{\Lambda^4}{f^2 m^{2}_{\mathrm{mono}}}.
\end{equation}
For $\kappa \ll 1$ we essentially have a quadratic potential whereas for $\kappa\gtrsim 1$ we have pronounced wiggles.
Examples are shown in Fig.~\ref{potsplot}.

Indeed the essential features of the classical evolution of a homogeneous field can be entirely characterized by $\kappa$ and the dimensionless expansion parameter
\begin{equation}
h=\frac{H}{m_{\mathrm{mono}}}.
\end{equation}
This can be seen by rescaling,
\begin{equation}
\label{eq:rescale}
t\rightarrow\tau \equiv m_{\mathrm{mono}}t, \quad \mathbf{x}\rightarrow \varix \equiv m_{\mathrm{mono}}\mathbf{x} \qquad \phi\rightarrow\varphi \equiv \frac{\phi}{f}.
\end{equation}
Using these dimensionless variables the equation of motion reads,
\begin{equation}
\ddot{\varphi}+3h\dot{\varphi}+\varphi+\kappa^2 \sin(\varphi)=0 \ ,
\end{equation}
where a dot $\dot{}$ now denotes a derivative w.r.t.~$\tau$.

\begin{figure} 
\centering
 \begin{overpic}[width=0.48\textwidth]{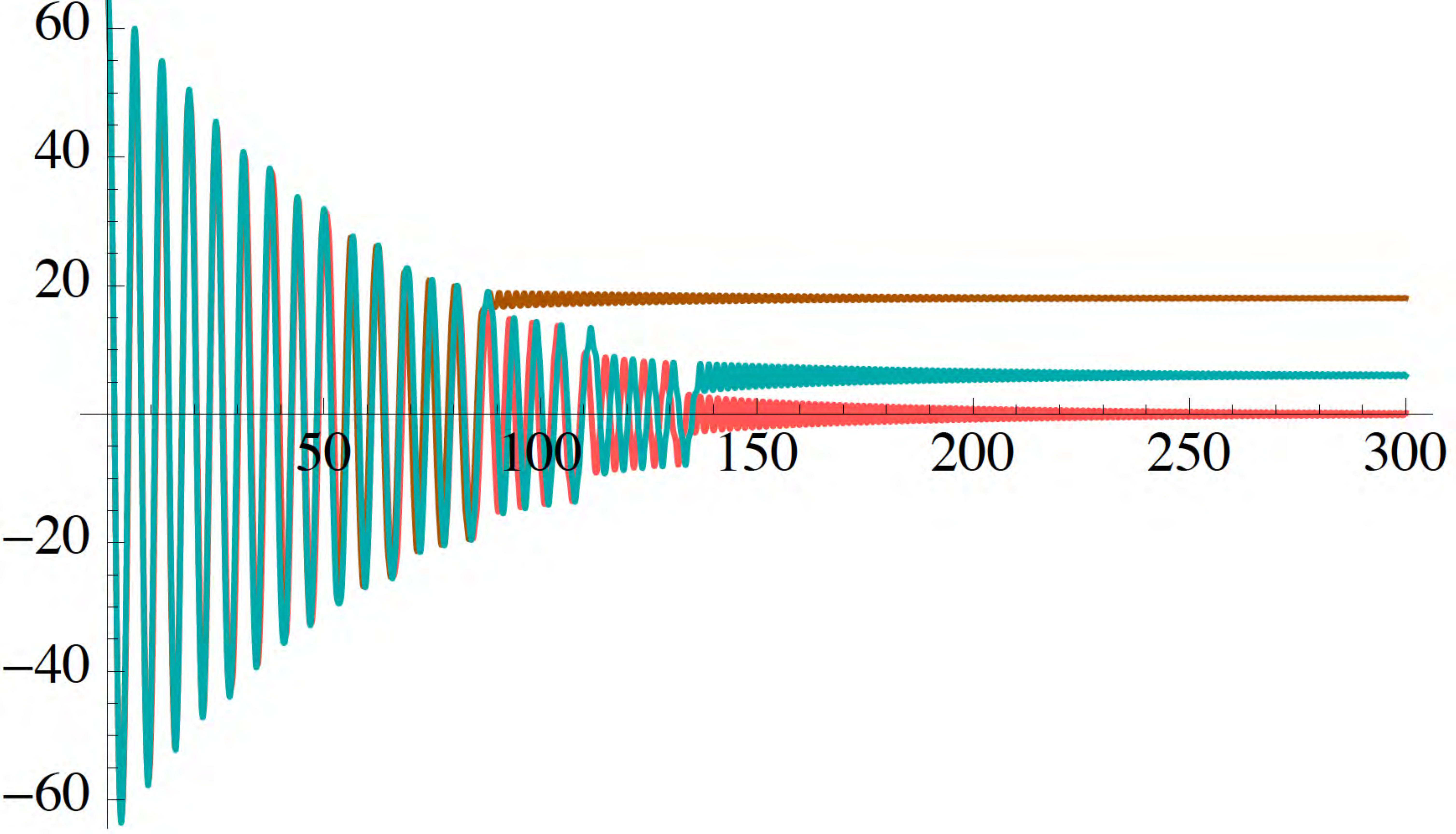}
 \put (14,54) {$\varphi$} \put (95,19) {$\tau$} 
\end{overpic}
  \caption{Time evolution of $\varphi$ vs.~$\tau$ for $h=0.01$, $\kappa=5.0$ and $\varphi_{\mathrm{initial}}=2\pi \cdot 10.3$ (red), $\varphi_{\mathrm{initial}}=2\pi \cdot 10.4$ (brown) and $\varphi_{\mathrm{initial}}=2\pi \cdot 10.5$ (cyan). The corresponding potential is shown in figure \ref{potsplot} (a,b). Note that $\varphi$ settles in different minima for only slightly different initial conditions.}	
\label{evolution1}
\end{figure}

\begin{figure}
    \centering
      \subfigure[
    ]
    {  \begin{overpic}[width=0.44\textwidth]{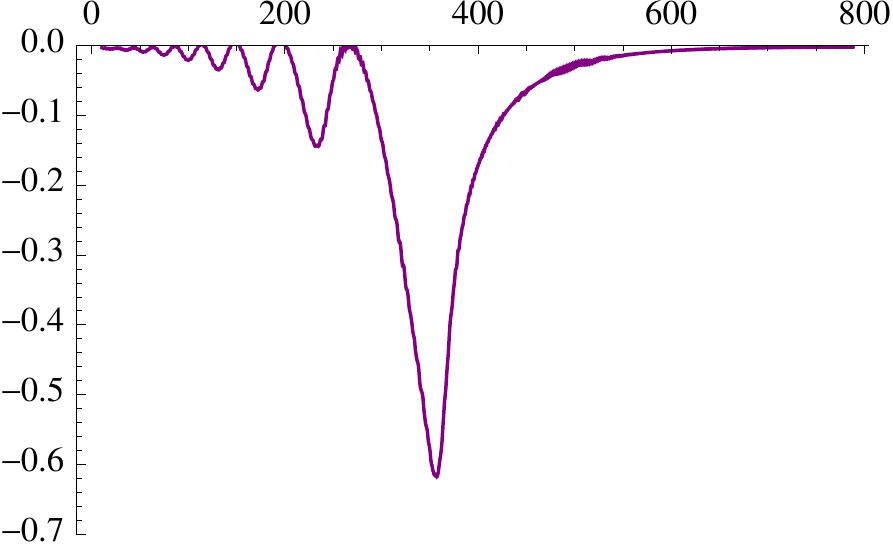}
 \put (46,64) {$\tau$} \put (13,2) {$w$} 
\end{overpic}
\label{fig:eosM2i10}
 }
\subfigure[
      ]
    {   \begin{overpic}[width=0.44\textwidth]{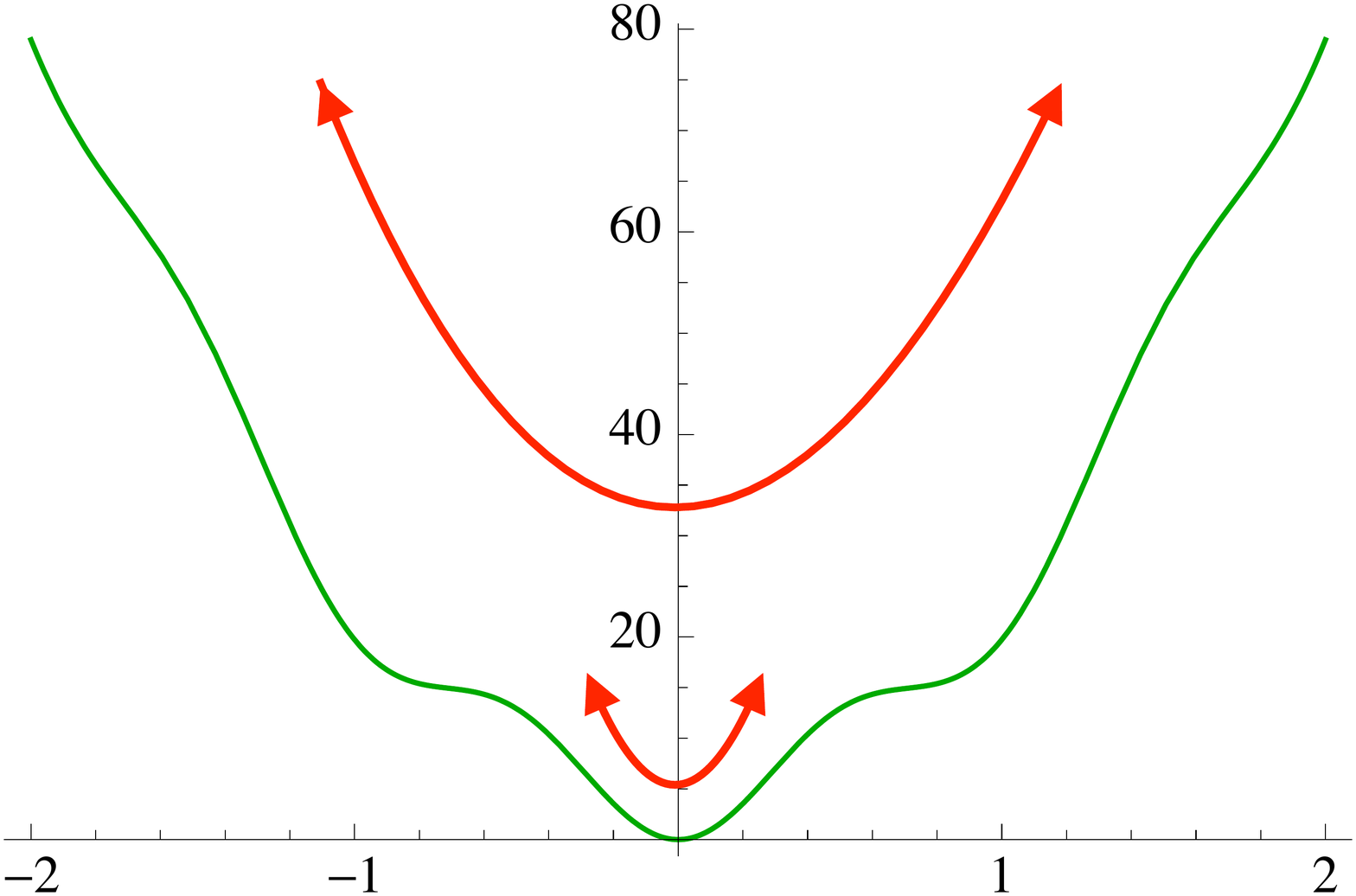}
 \put (52,64) {$\tfrac{V}{f^2 m_{\mathrm{mono}}^2}$} \put (86,12) {$\varphi/2 \pi$} 
\end{overpic}
\label{fig:VM2}
 }
\caption{(a): Equation of state parameter $w$ vs.~$\tau$ for $h=0.005$, $\kappa=2.0$ and $\varphi_{\mathrm{initial}}=2\pi \cdot 10.25$. (b): The corresponding potential for $\kappa=2.0$.}
\label{fig:eos}
\end{figure}

As discussed in the previous section we are particularly interested in the case of large initial field values $\phi\gg f\leftrightarrow \varphi\gg 1$, which will allow us to increase the dark matter density and enter previously unachievable parameter regions. Let us first investigate the behaviour in this region. It is clear that for $\kappa\to 0$ the evolution will be that of a simple massive field with mass $m_{\mathrm{mono}}$. However, as we can see from Fig.~\ref{potsplot}, even for sizeable non-vanishing $\kappa$ its effect on the potential is very small for sufficiently large field values of $\varphi$. 

One quantity that will be affected by the presence of the $\cos$-term could be the overall oscillation frequency. In absence of the $\cos$ contribution the angular frequency is $\Omega_0=1$ (in units of $m_{\mathrm{mono}}$). In Fig.~\ref{fig:frequency} we plot the (angular) oscillation frequency $\Omega$ (for $h=0$) as a function of the initial field value and for the values of $\kappa=2.0$ and $\kappa=6.0$. We can clearly see that for large initial field values the effect of the $\cos$-term is negligible and the angular frequency approaches the value $\Omega_0=1$.

However, as the Universe expands the oscillations are damped and the field values become smaller. So we will never entirely stay in a regime with large field values.
In the small field value regime the effect of the $\cos$-term is more profound.
In particular, for $\kappa\gtrsim 1$ the potential can have additional local minima.
This can obviously have significant effects on the cosmological evolution. Depending on the initial conditions the
field can then get stuck in different local minima. 
This is shown in fig.~\ref{evolution1} where we display three  solutions of the equation of motion with slightly different initial conditions that end up in different minima.

For cosmology, a more important quantity is the equation of state, or the dilution as the Universe expands.
In fig.~\ref{fig:eosM2i10} we plot the initial value for the equation of state parameter (averaged over a period of oscillation and for $h\ll1$) as a function of time. Initially, when the field amplitude is large, the field is oscillating in the broad parabola with curvature $m_{\mathrm{mono}}^2$, as indicated by the wide red arrow in figure \ref{fig:VM2}. The equation of state is then close to $w=0$, i.e.~the equation of state for dark matter. Similarly, at late times it is again close to $w=0$ when the field is oscillating about the minimum at $\varphi=0$, as shown by the narow red arow in figure \ref{fig:VM2}). However, in fig.~\ref{fig:eosM2i10} we can also see that there is a time when the equation of state is quite different from $w=0$. This is typically the case when the field runs relatively slowly through plateau-like regions of the potential. For the example in figure  \ref{fig:VM2} such plateau-like regions exist for $\tfrac{\varphi}{2 \pi} \in \pm [0.5, 1]$. During the slow field evolution on a plateau the energy density is dominated by the near-constant potential energy of the field $\varphi$ giving rise to a negative equation of state.

If such deviations from $w=0$ occur early, before matter radiation equality, they should be relatively unconstrained.
If they happen later one expects deviations in the cosmological evolution and in particular structure formation (cf, e.g.~\cite{Das:2006ht,Arias:2012az}). We leave a detailed investigation of the cosmological constraints to future work.

\section{Growth of (quantum) fluctuations}\label{quantum}
As we have seen in the previous section, for non-vanishing values of $\kappa$ the classical evolution can be altered significantly. However, additional and perhaps more drastic changes can also happen on the quantum level.

As discussed above, for $\kappa\gtrsim 1$ we can have additional minima and maxima in the potential. At a maximum, however, the curvature of the potential is negative, which can be associated with a tachyonic instability on the quantum level. This indicates that fluctuations, including quantum fluctuations, may grow and can in some cases become important. 

In the following we perform an analysis of the linearized equations of motion for the fluctuations (for $h=0$). Indeed, the system exhibits a behaviour similar to parametric resonance where in particular fluctuations of a certain size (or equivalently momentum) are growing exponentially.  On the particle level the growth of fluctuations can be interpreted as production of particles with non-vanishing momenta. As these momenta are typically semi-relativistic (or even relativistic) this will lead to potentially observable effects in the cosmological evolution.

A full discussion, in particular of the non-linear regime, is beyond the scope of the present paper. Nevertheless, we provide some estimates on parameter regimes where the growth of fluctuations and the corresponding particle production become relevant. 

\subsection{Growing fluctuations from the linearized equations of motion}
Here we want to show that fluctuations indeed grow and quantify this effect. To get a first impression of the qualitative behaviour we consider the situation without expansion, i.e.~$h=0$. We will return to a more complete discussion including expansion in future work.

Initially we can assume that the fluctuations are small.
Therefore we will use the linearized equations of motion for the field fluctuations,
\begin{equation}
\varphi(\tau, \varix)=\varphi_{0}(\tau)+\delta \varphi(\tau, \varix),
\end{equation}
where $\varphi_{0}(\tau)$ denotes the homogeneous background solution discussed in section \ref{classical}. It will be useful to expand the fluctuation in Fourier modes as
\begin{equation}
\delta \varphi(\tau, \varix) = \int \frac{d^3 k}{(2 \pi)^{3/2}} \ c_{\mathbf{k}}(\tau)\exp(i\mathbf{k}\cdot\varix) \ .
\end{equation}
Note that we have rescaled from $\mathbf{x}$ to the dimensionless variable $\varix$ in \eqref{eq:rescale}. Hence $\mathbf{k}$ above is measured in units of $m_{\mathrm{mono}}$, i.e.~the physical momentum is given by $\mathbf{k}_{phys}=m_{\mathrm{mono}} \mathbf{k}$.
As $\delta \varphi$ is real we require $c_{-\mathbf{k}}= c_{\mathbf{k}}^*$. To linear order the equation of motion for the modes $c_{\mathbf{k}}$ is
\begin{equation}
\label{fluceom}
\ddot{c}_{\mathbf{k}} + \left(1 + k^2 - \kappa^2 \cos(\varphi_0) \right) c_{\mathbf{k}}=0 \ ,
\end{equation}
where we also defined $k \equiv |\mathbf{k}|$.
It is the dependence on the background solution $\varphi_{0}(\tau)$ which makes this equation hard to solve. 

In fact, equation \eqref{fluceom} is of Hill type, i.e.~it is a differential equation of the form $\ddot{c} + p(\tau) c =0$ where $p(\tau)$ is a periodic function. According to Floquet's theorem one can always find two solutions to Hill's equation of the form $e^{\eta \tau} F(\tau)$ and $e^{-\eta \tau} F(-\tau)$ where $F(\tau)$ is periodic. Depending on $p(\tau)$ the parameter $\mu$ can be either imaginary, real or complex. Whenever $\mu$ has a real part there is an instability as one of the solutions exhibits exponential growth. No growth occurs if $\mu$ is found to be purely imaginary. For our system this can also be understood as follows. While we expect rapid growth of fluctuations due to the presence of tachyonic instabilities, these instabilities are only present part of the time. Growth of fluctuations can only occur if resonance conditions are met, i.e.~growth will only occur for certain values of $k$.\footnote{This is closely related to the phenomenon of parametric resonance that has been widely discussed in the context of reheating (for early work see \cite{hep-th/9405187, hep-ph/9704452} and \cite{1001.2600} for a review). For a somewhat different example of large growth of fluctuations in the context of axion-like particles see e.g.~\cite{1508.04136}.}

Furthermore, as eq.~\eqref{fluceom} is a second order differential equation we need two initial conditions.
In the following subsection~\ref{quantumini} we will discuss those suggested by (static) quantum theory. In principle however, the initial conditions depend on the previous evolution of the Universe, e.g.~inflation. In that sense the discussion in subsection~\ref{quantumini} should be considered as a somewhat simplistic example.

More important is, however, the fact that fluctuations can grow rather rapidly. Due to the linearity of eq.~\eqref{fluceom}
this growth applies to essentially any initial condition (except for fine-tuned initial conditions) and is in that sense generic.
This is what we will investigate in the remainder of this subsection. 

In the following, we solve eq.~\eqref{fluceom} numerically and present the results. Our strategy is as follows. First we obtain the homogeneous solution for a given initial value $\varphi_{\mathrm{initial}}$ (for simplicity we take $\dot{\varphi}_{\rm initial}=0$).\footnote{This is sufficient to obtain the growth exponent.} Then we insert this solution into Eq.~\eqref{fluceom} and solve numerically for a significant range of $k$. From this solution we can then extract the growth exponent $\eta(k)$. To check our results we also calculated the growth exponent using a different approach: it can also be determined using Hill's determinental method as described in appendix \ref{sec:Hill}. We checked that both methods give the same results.

\subsection*{Results}

\begin{figure}
    \centering
      \subfigure[
    ]
    {   \includegraphics[width=0.44\textwidth]{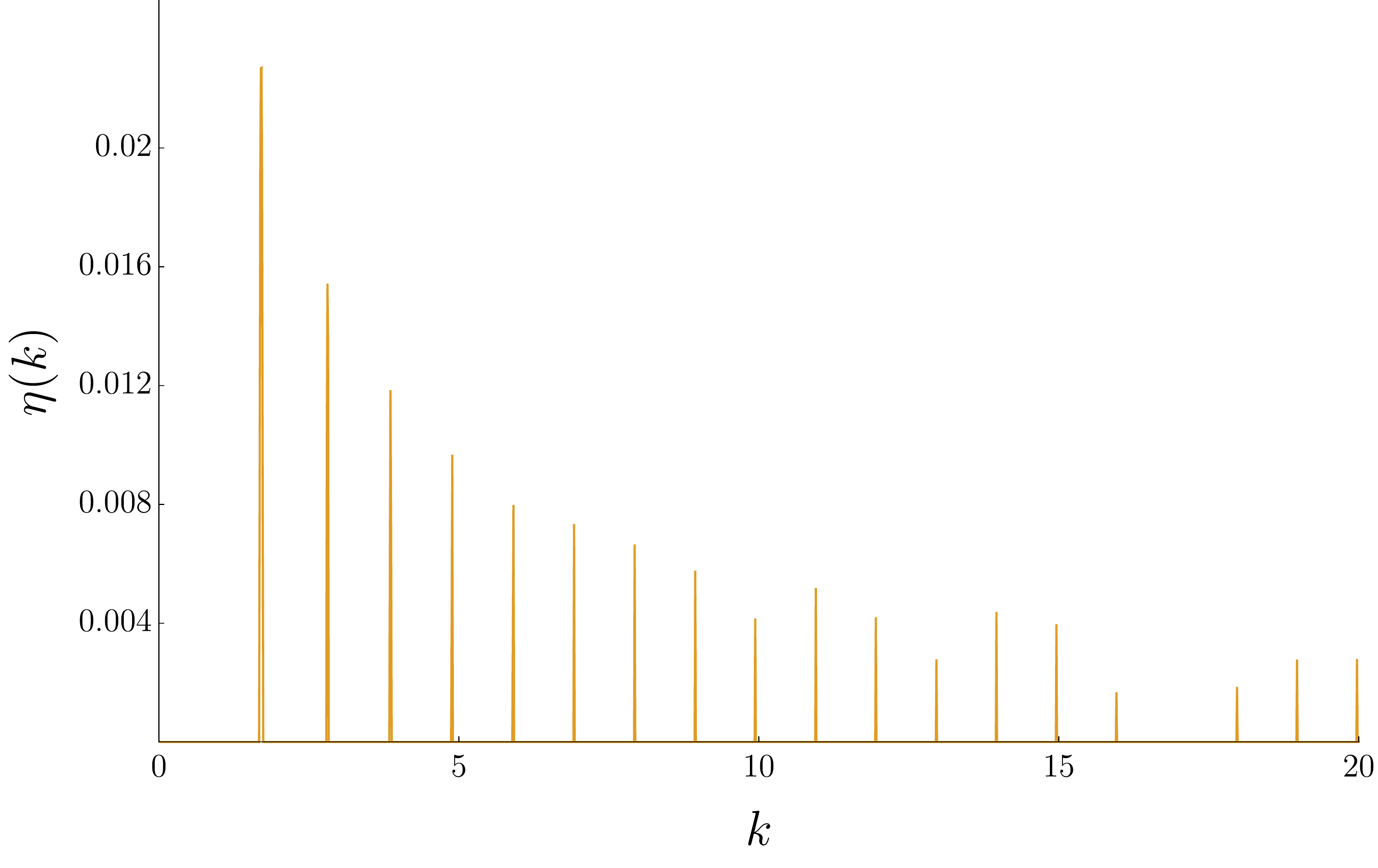}
   \label{fig:m2muk20}
 }
  \subfigure[
    ]
    {   \includegraphics[width=0.44\textwidth]{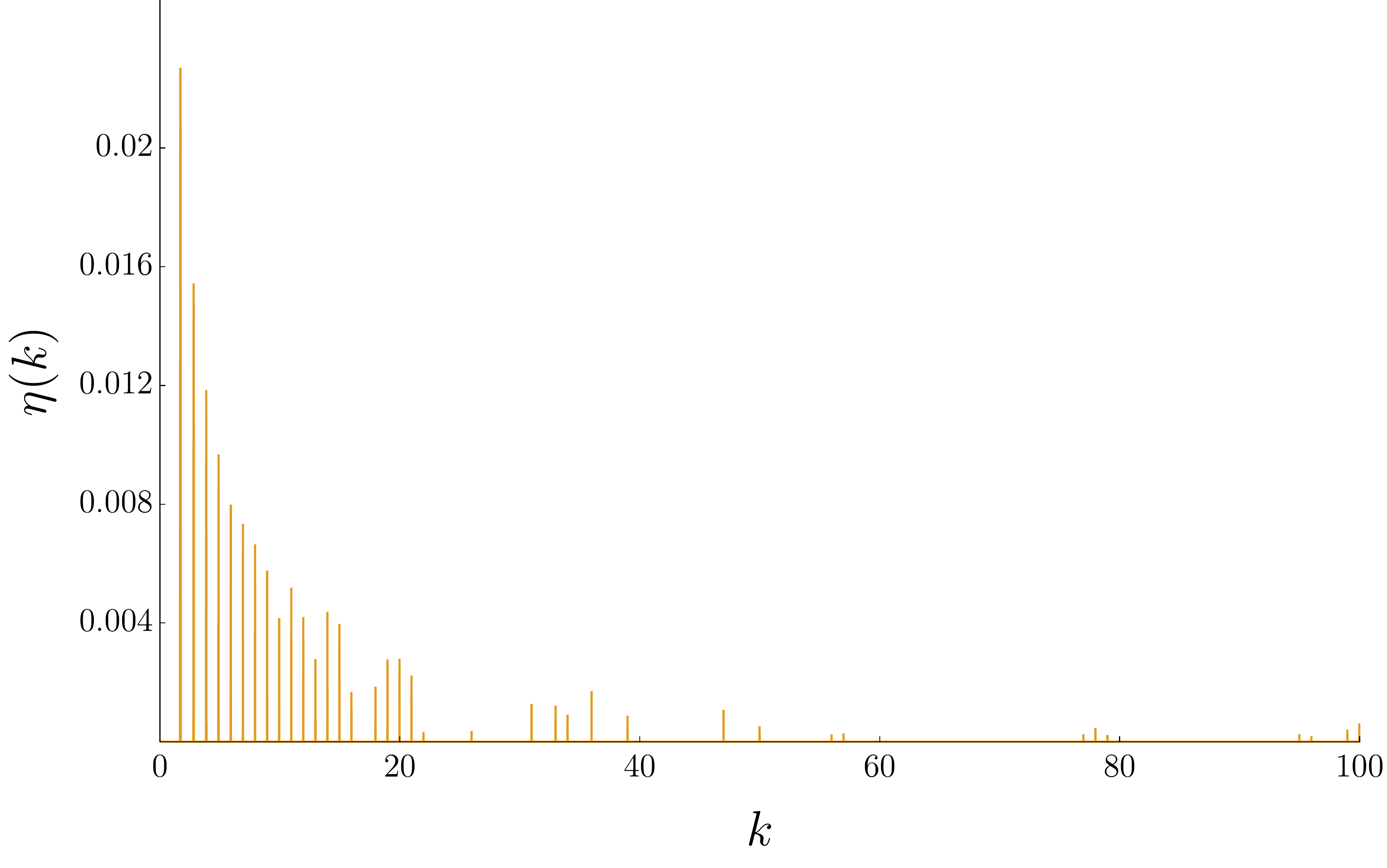}
   \label{fig:m2muk100}
 }
 \\
      \subfigure[
    ]
    {   \includegraphics[width=0.44\textwidth]{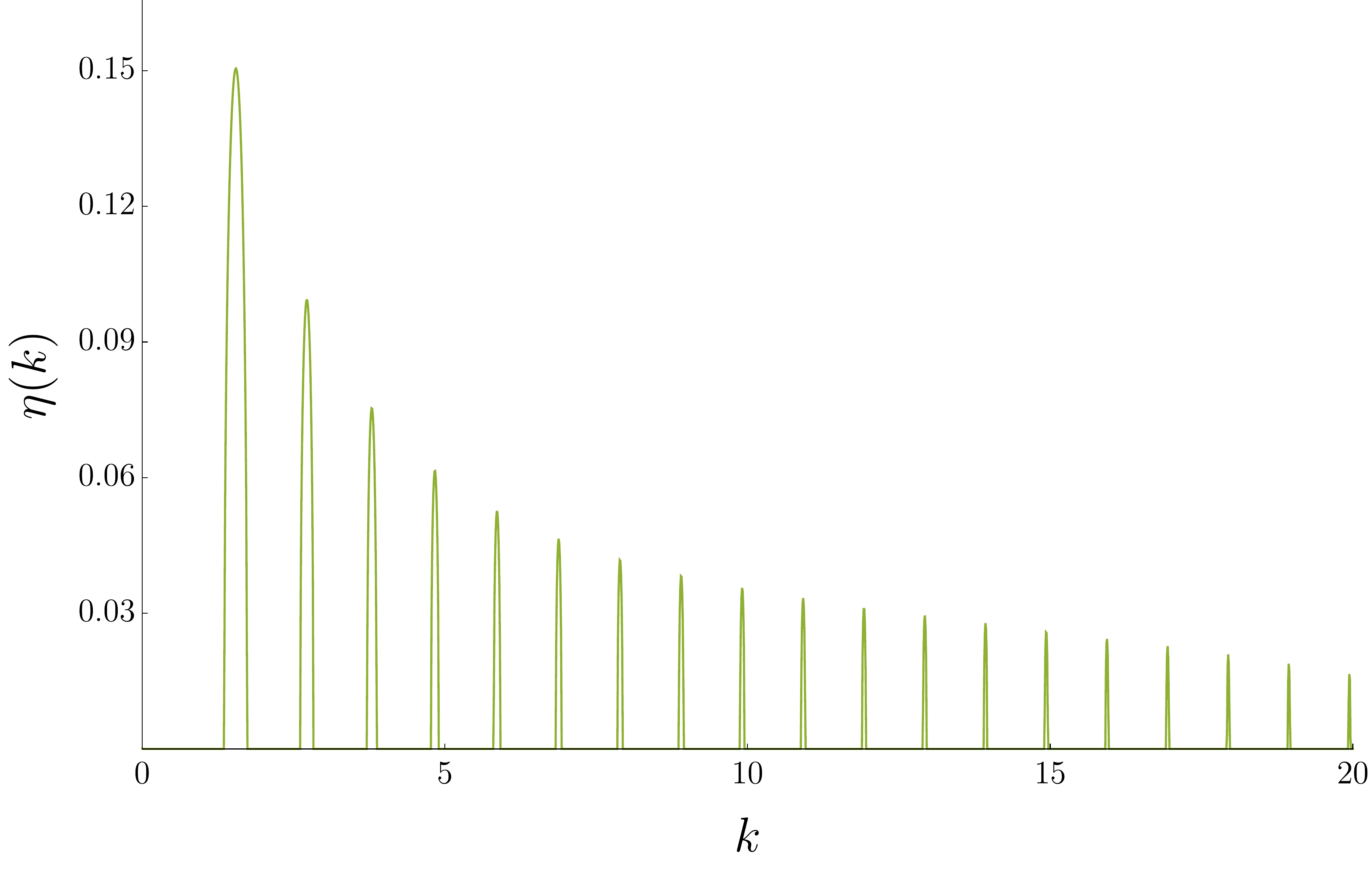}
   \label{fig:m5muk20}
 }
 \subfigure[
   ]
    {   \includegraphics[width=0.44\textwidth]{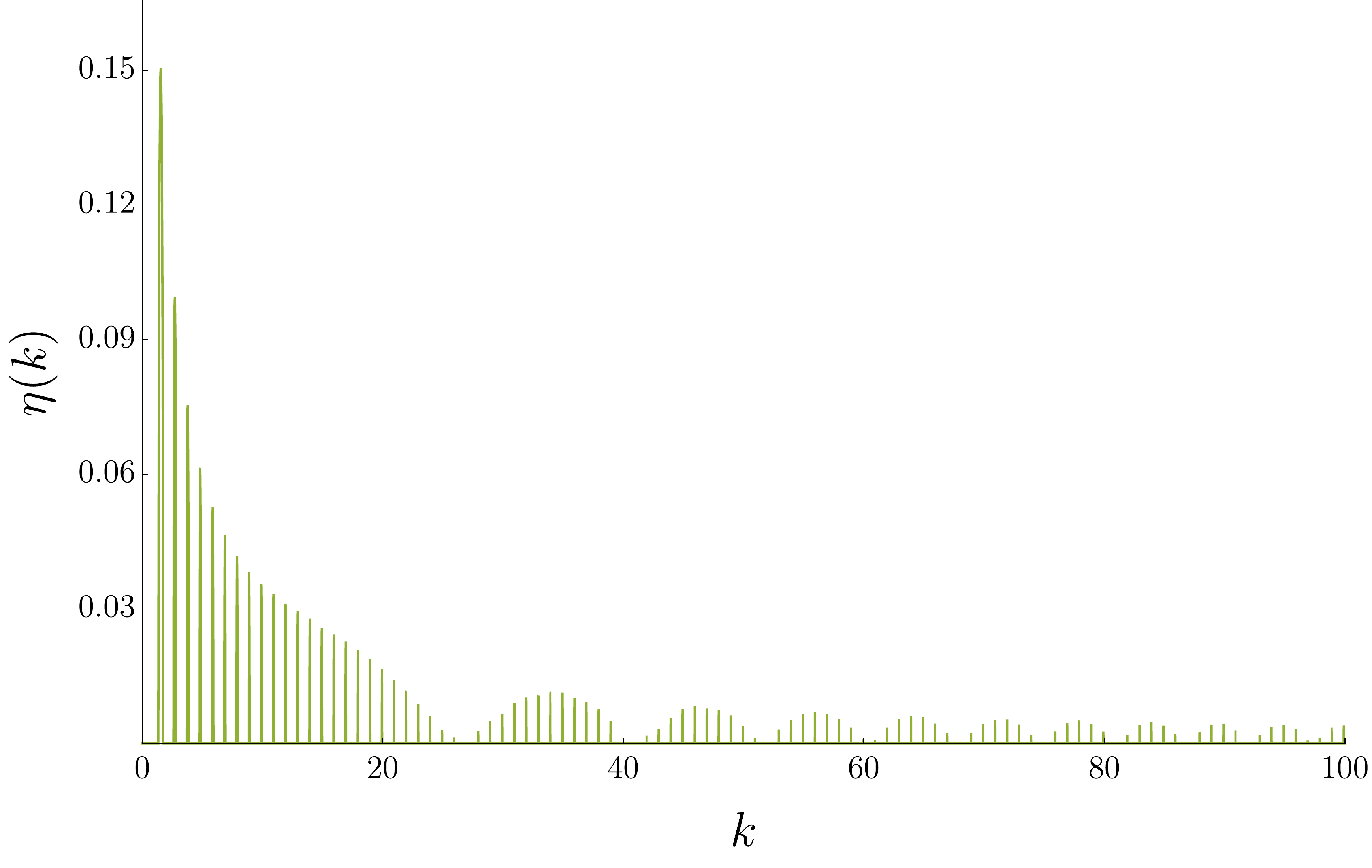}
   \label{fig:m5muk100}
 }
 \\
     \subfigure[
    ]
    {   \includegraphics[width=0.44\textwidth]{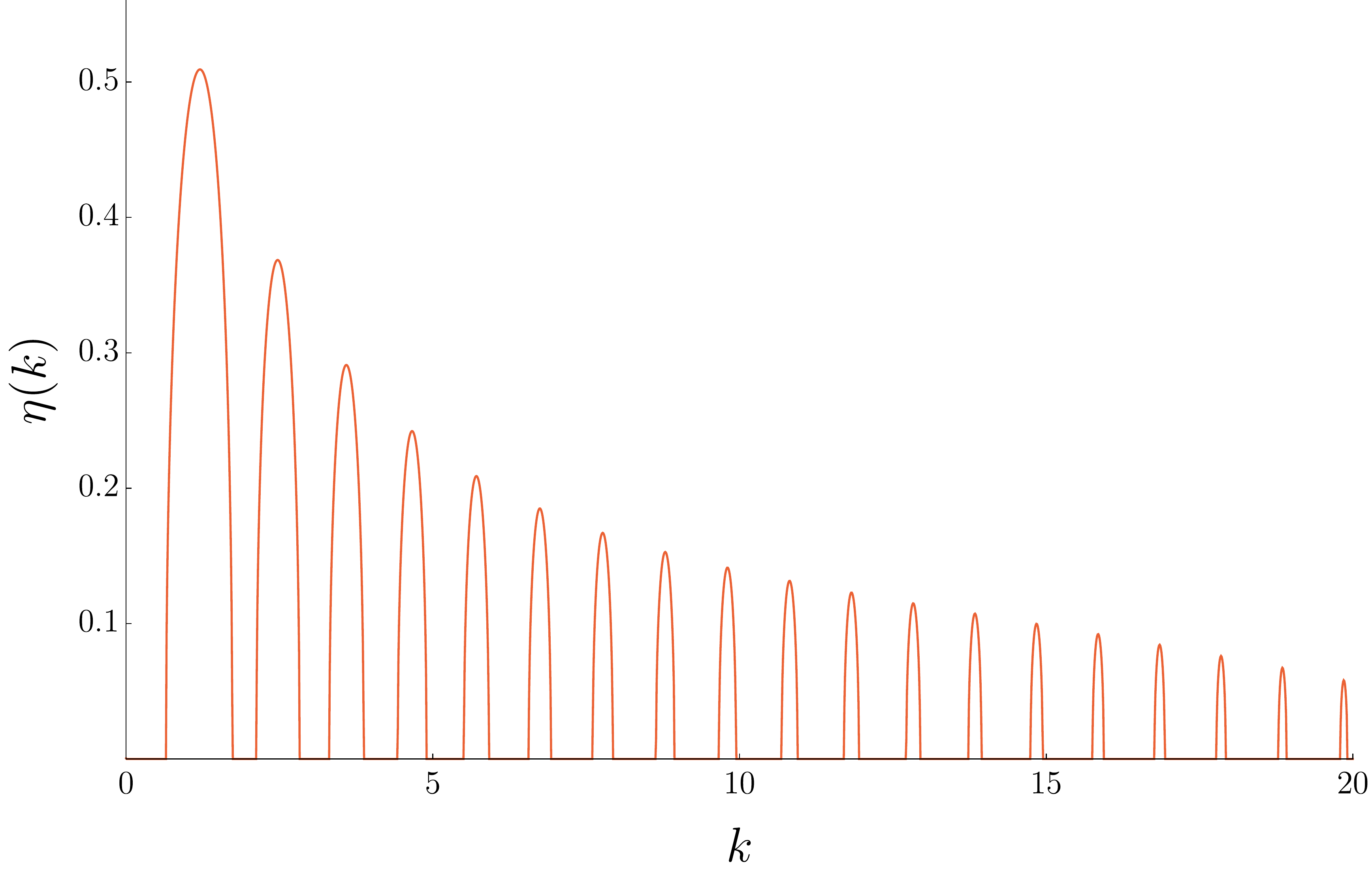}
   \label{fig:m10muk20}
 }
   \subfigure[
   ]
    {   \includegraphics[width=0.44\textwidth]{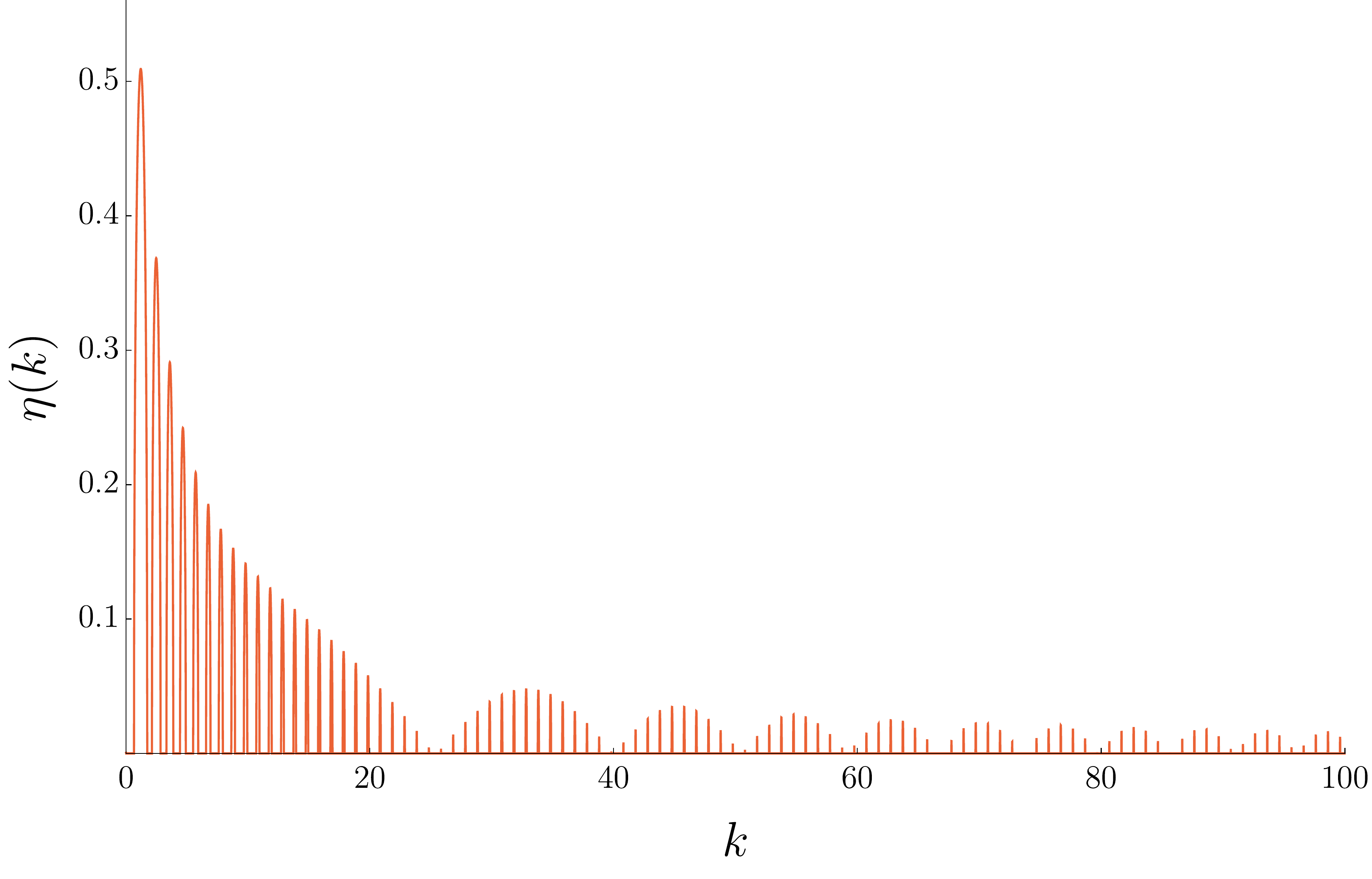}
   \label{fig:m10muk100}
 }
\\ 
\subfigure[
    ]
    {   \includegraphics[width=0.44\textwidth]{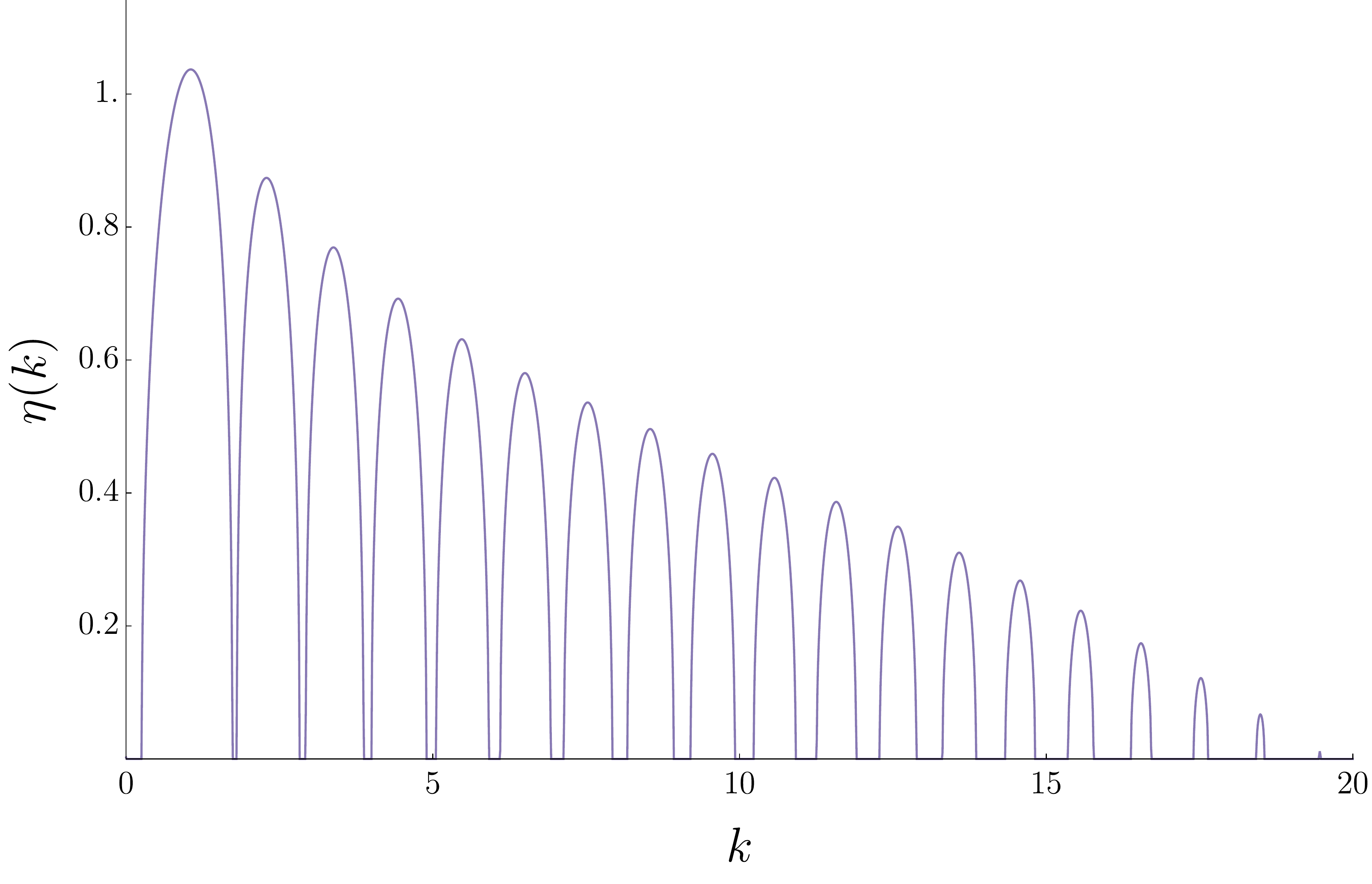}
   \label{fig:m20muk20}
 }
 \subfigure[
    ]
    {   \includegraphics[width=0.44\textwidth]{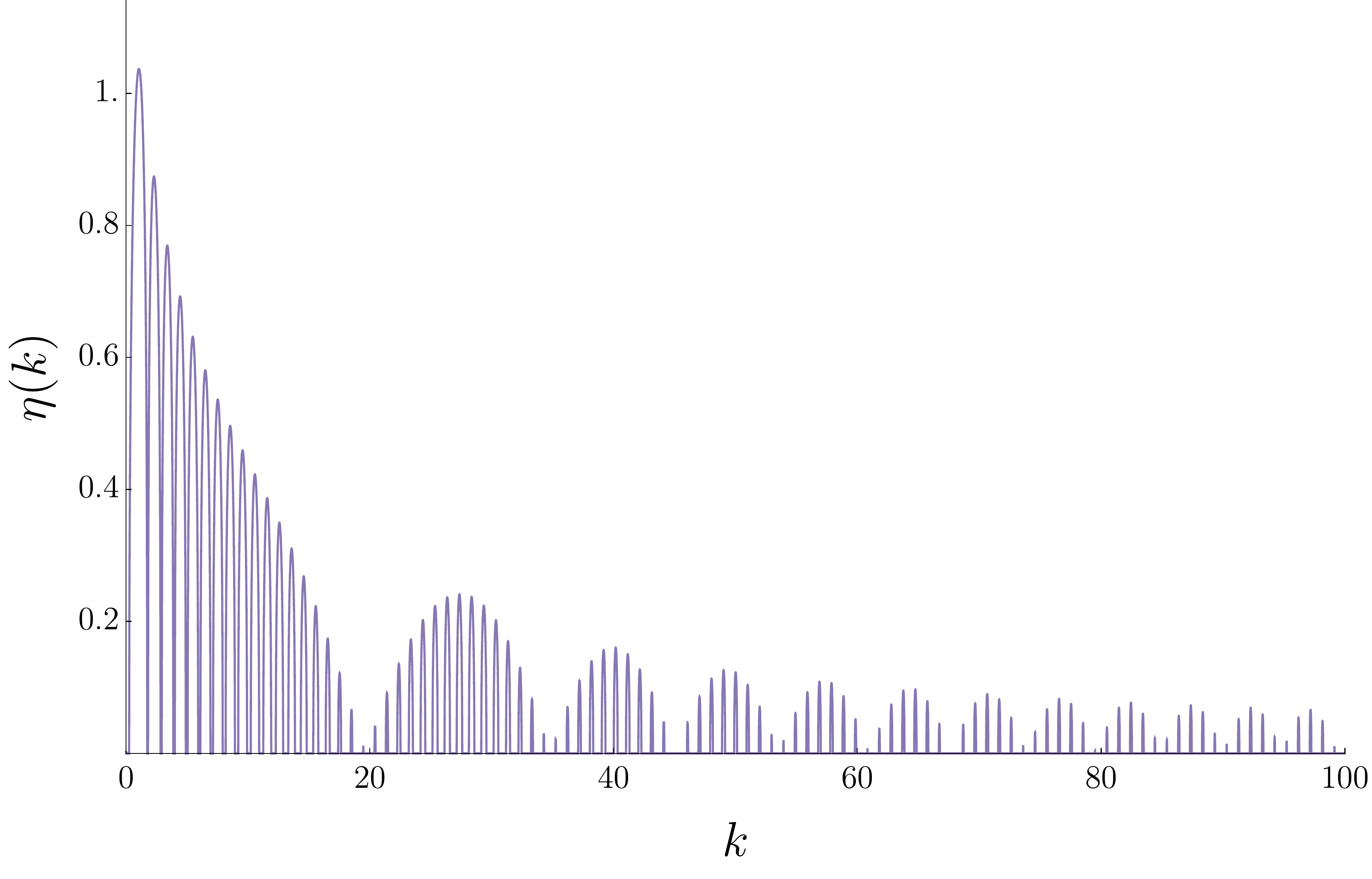}
   \label{fig:m20muk100}
 }
 \caption{Plots of $\eta(k)$ vs.~$k$ for $\tfrac{\varphi_i}{2\pi}=100$ and (a,b) $\kappa=2.0$, (c,d) $\kappa=5.0$, (e,f) $\kappa=10$, (g,h) $\kappa=20$.}
    \label{fig:mrangemuplots}
  \end{figure}

\begin{figure}
    \centering
      \subfigure[
    ]
    {   \includegraphics[width=0.44\textwidth]{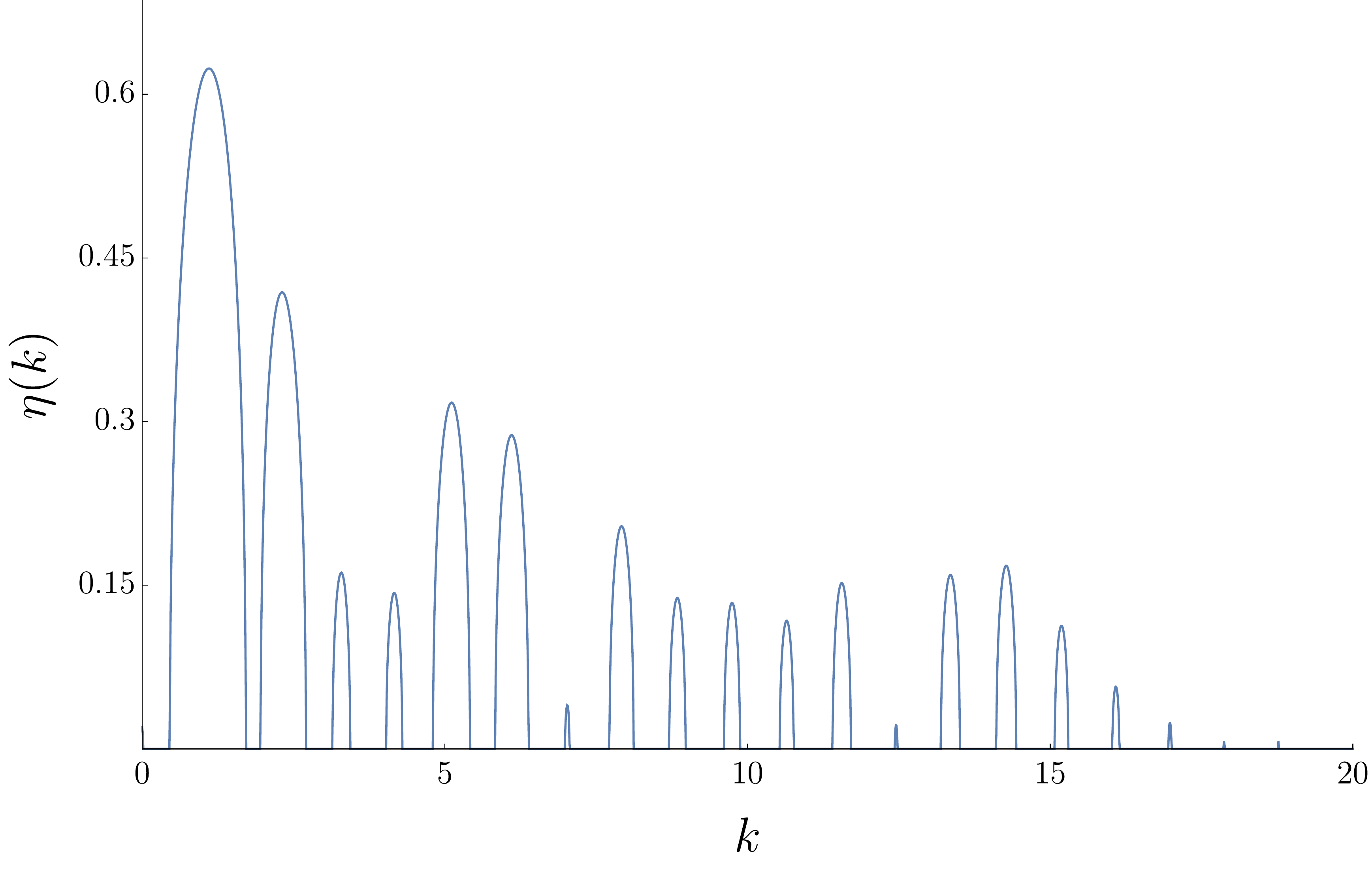}
   \label{fig:phi5muk20}
 }
  \subfigure[
    ]
    {   \includegraphics[width=0.44\textwidth]{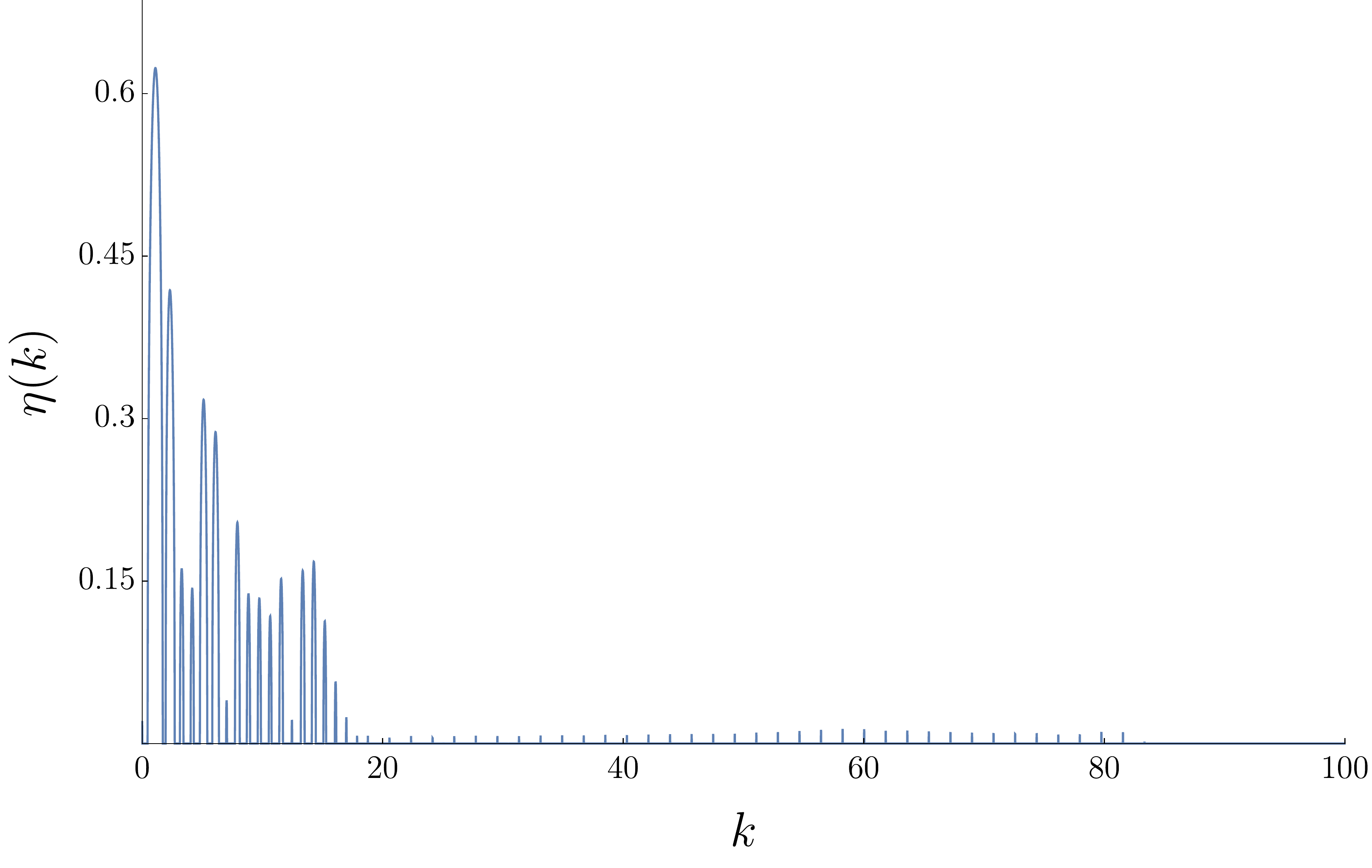}
   \label{fig:phi5muk100}
 }
 \\
      \subfigure[
    ]
    {   \includegraphics[width=0.44\textwidth]{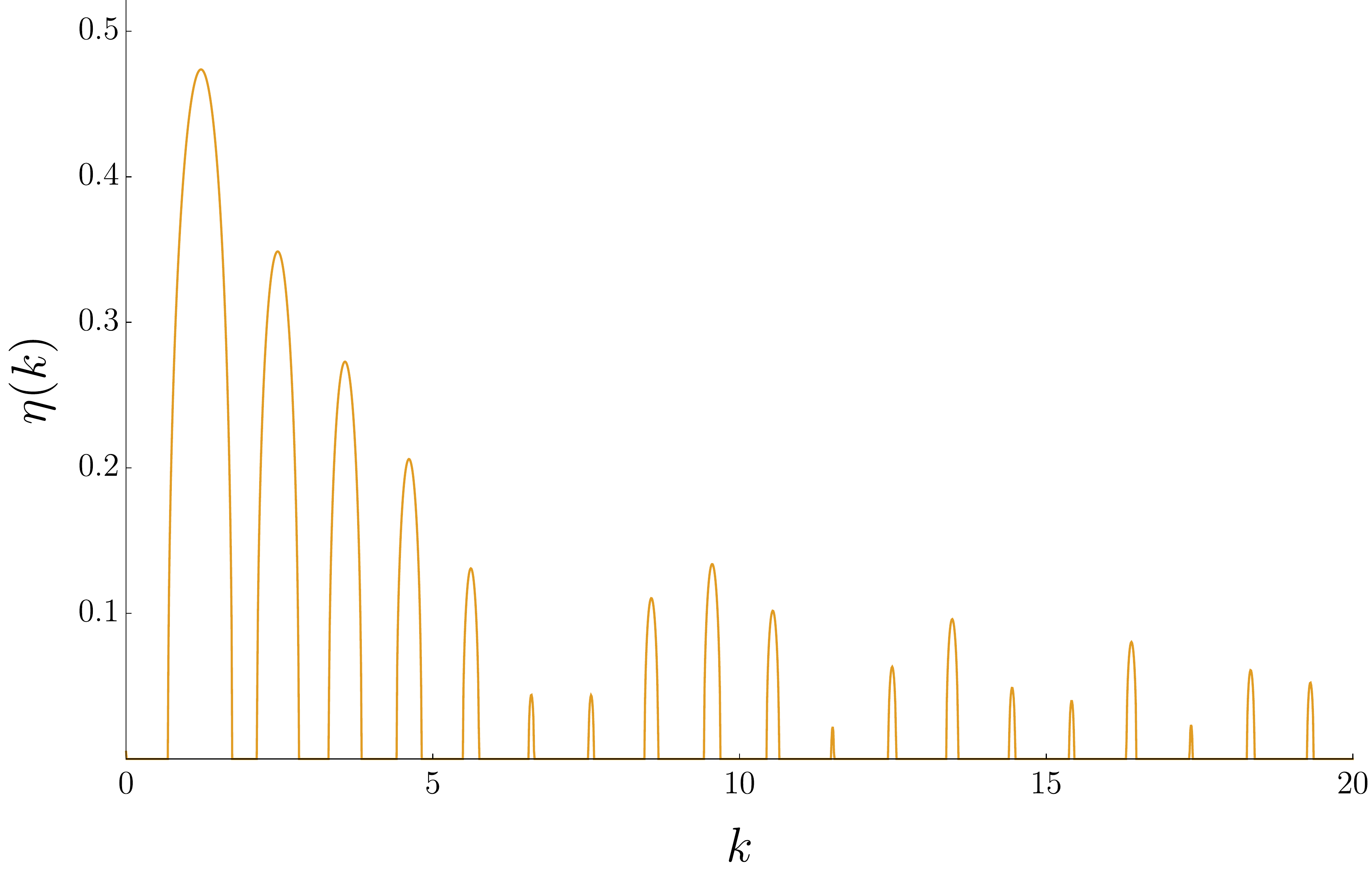}
   \label{fig:phi10muk20}
 }
 \subfigure[
   ]
    {   \includegraphics[width=0.44\textwidth]{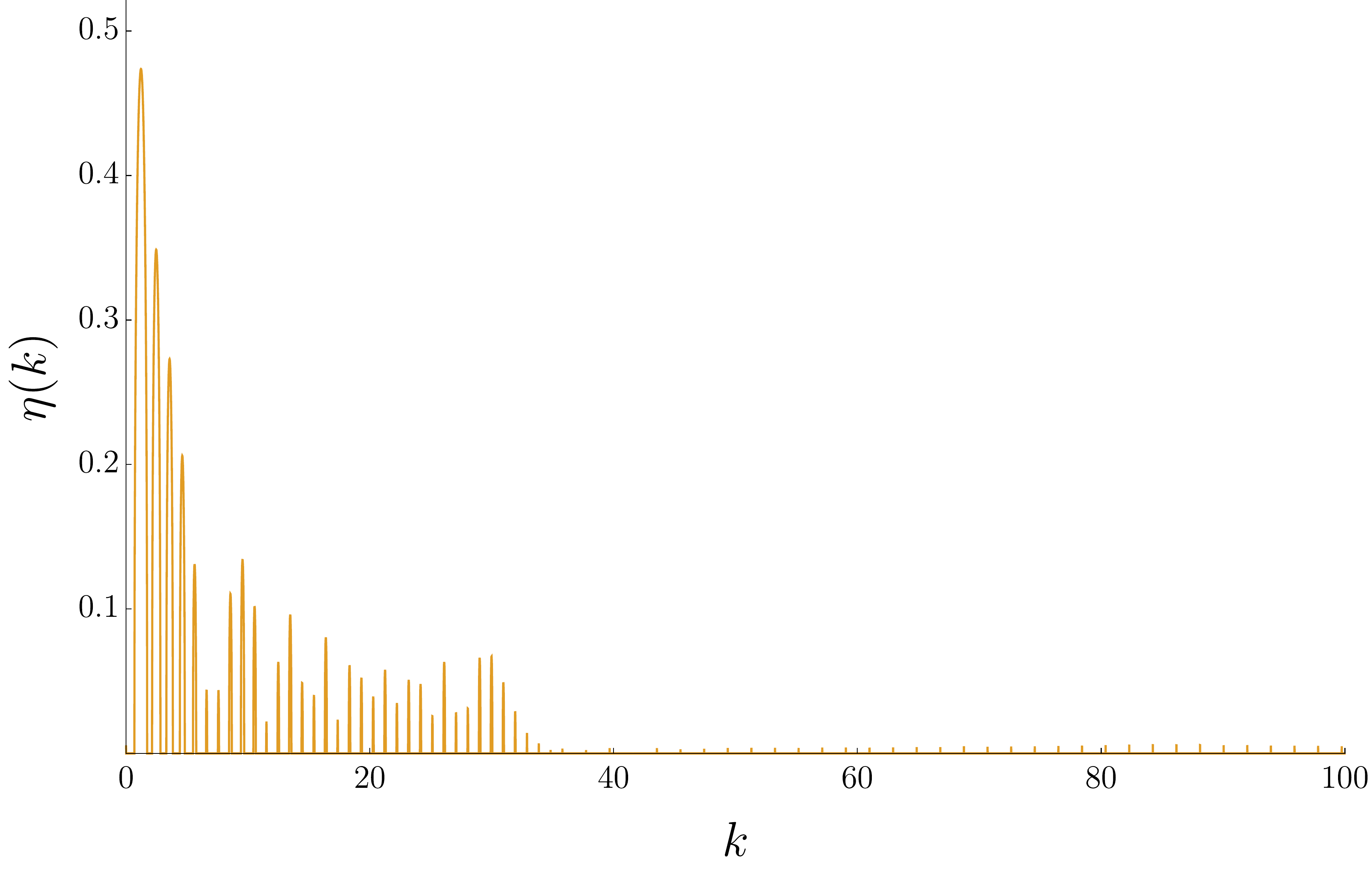}
   \label{fig:phi10muk100}
 }
 \\
     \subfigure[
    ]
    {   \includegraphics[width=0.44\textwidth]{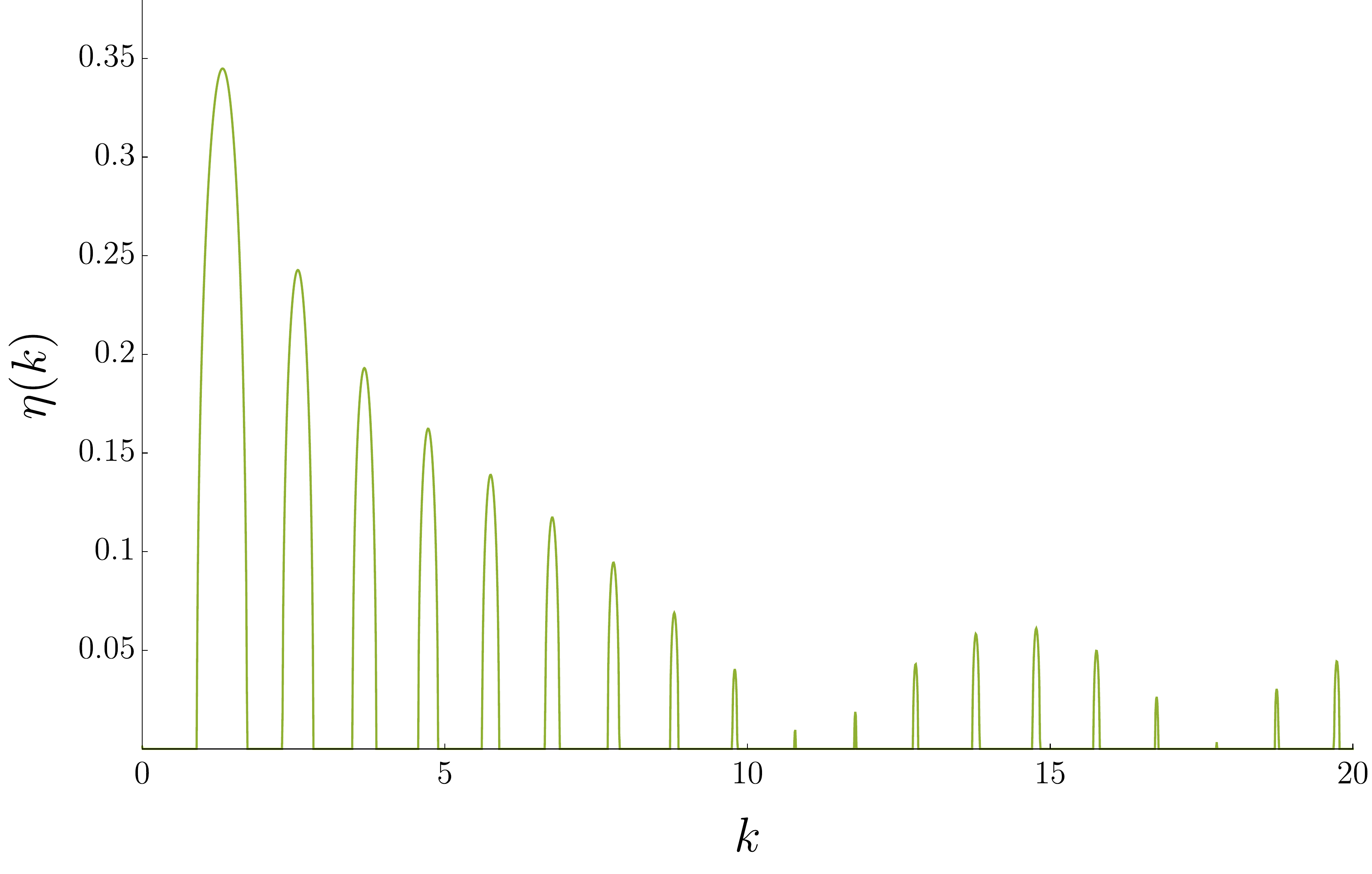}
   \label{fig:phi20muk20}
 }
   \subfigure[
   ]
    {   \includegraphics[width=0.44\textwidth]{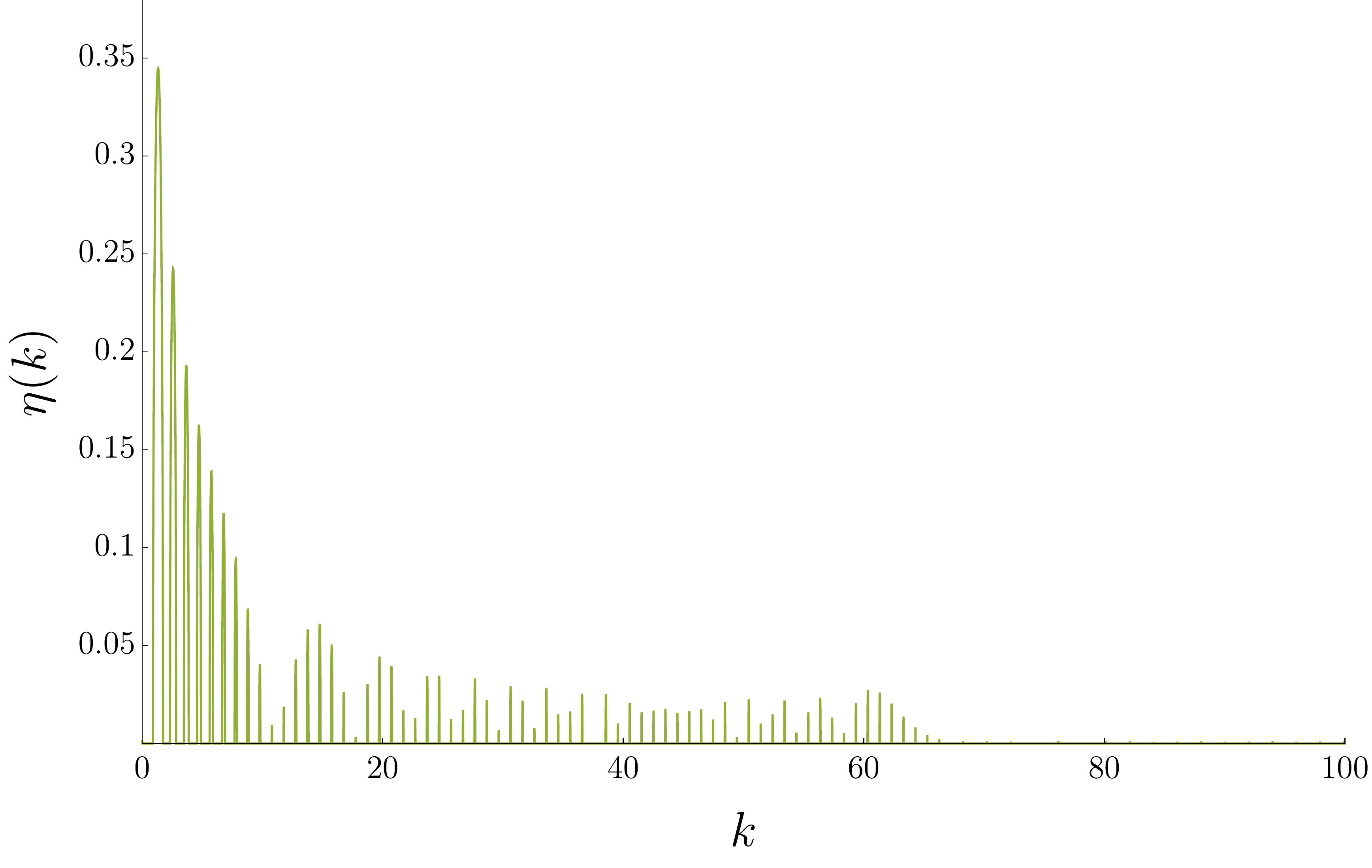}
   \label{fig:phi20muk100}
 }
\\ 
\subfigure[
    ]
    {   \includegraphics[width=0.44\textwidth]{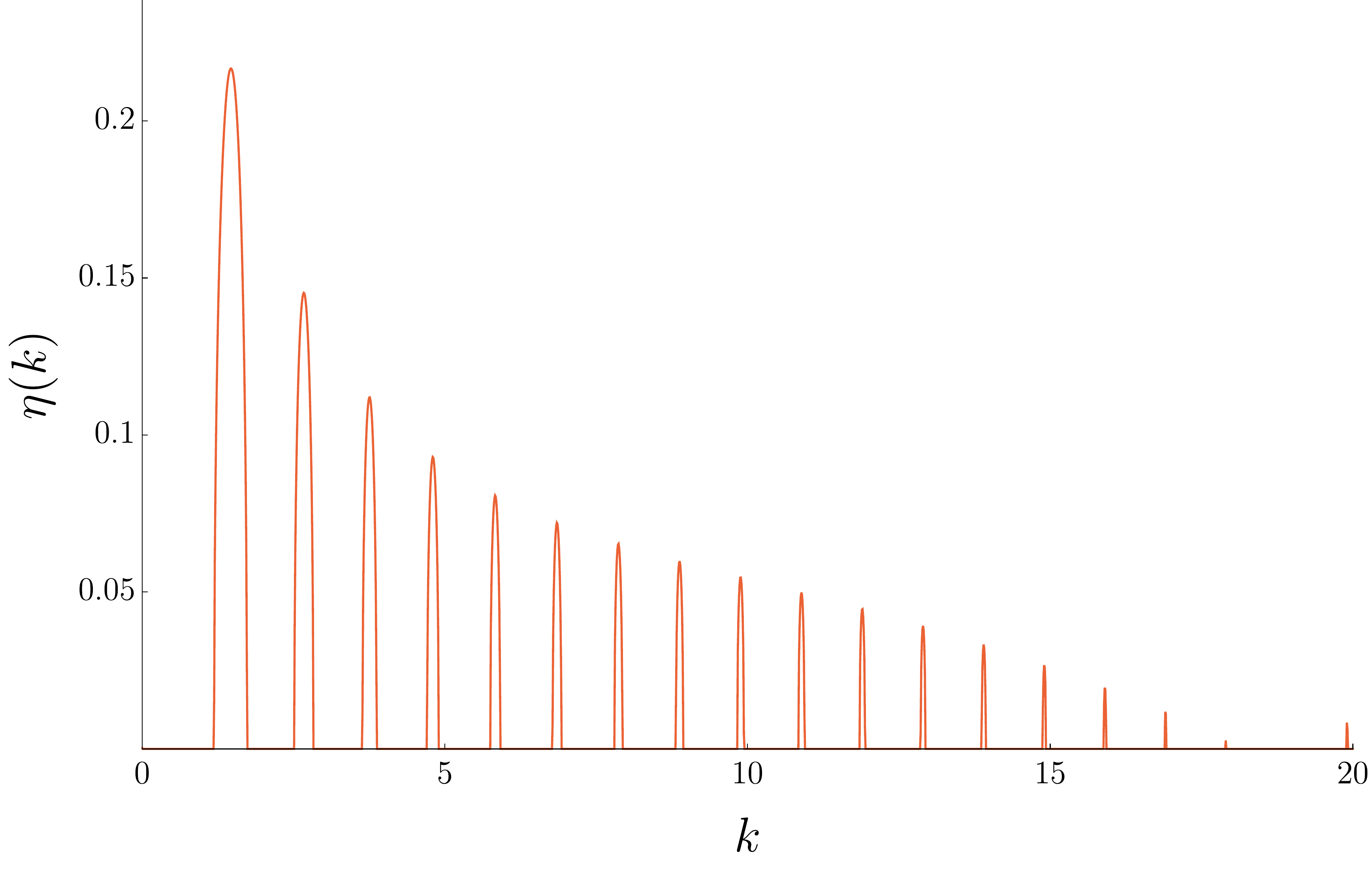}
   \label{fig:phi50muk20}
 }
 \subfigure[
    ]
    {   \includegraphics[width=0.44\textwidth]{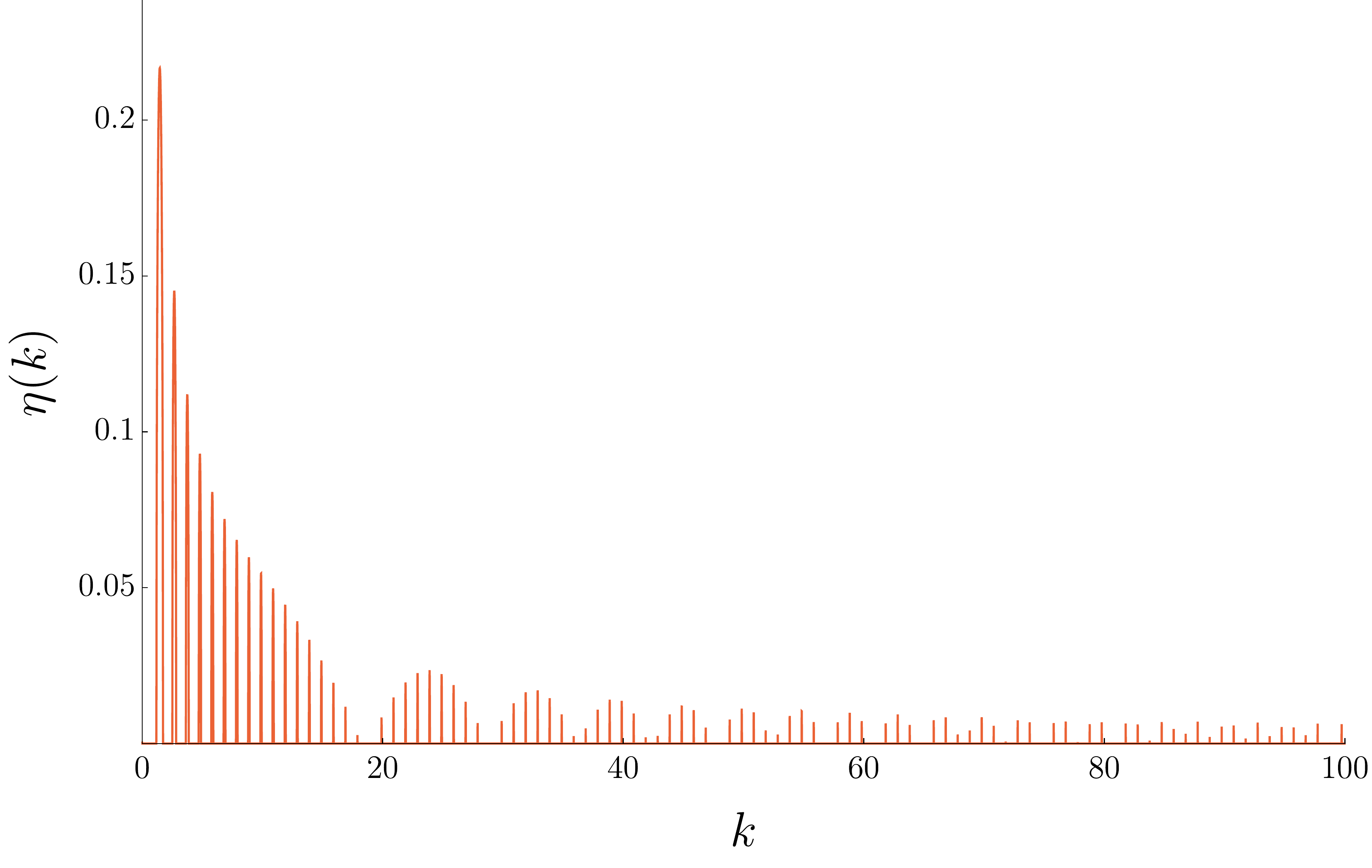}
   \label{fig:phi50muk100}
 }
      \caption{Plots of $\eta(k)$ vs.~$k$ for $\kappa=5.0$ and (a,b) $\tfrac{\varphi_{\mathrm{initial}}}{2\pi}=5.0$, (c,d) $\tfrac{\varphi_{\mathrm{initial}}}{2\pi}=10$, (e,f) $\tfrac{\varphi_{\mathrm{initial}}}{2\pi}=20$, (g,h) $\tfrac{\varphi_{\mathrm{initial}}}{2\pi}=50$.}
    \label{fig:phirangemuplots}
  \end{figure}

\begin{figure}
    \centering
      \subfigure[
    ]
    {   \includegraphics[width=0.48\textwidth]{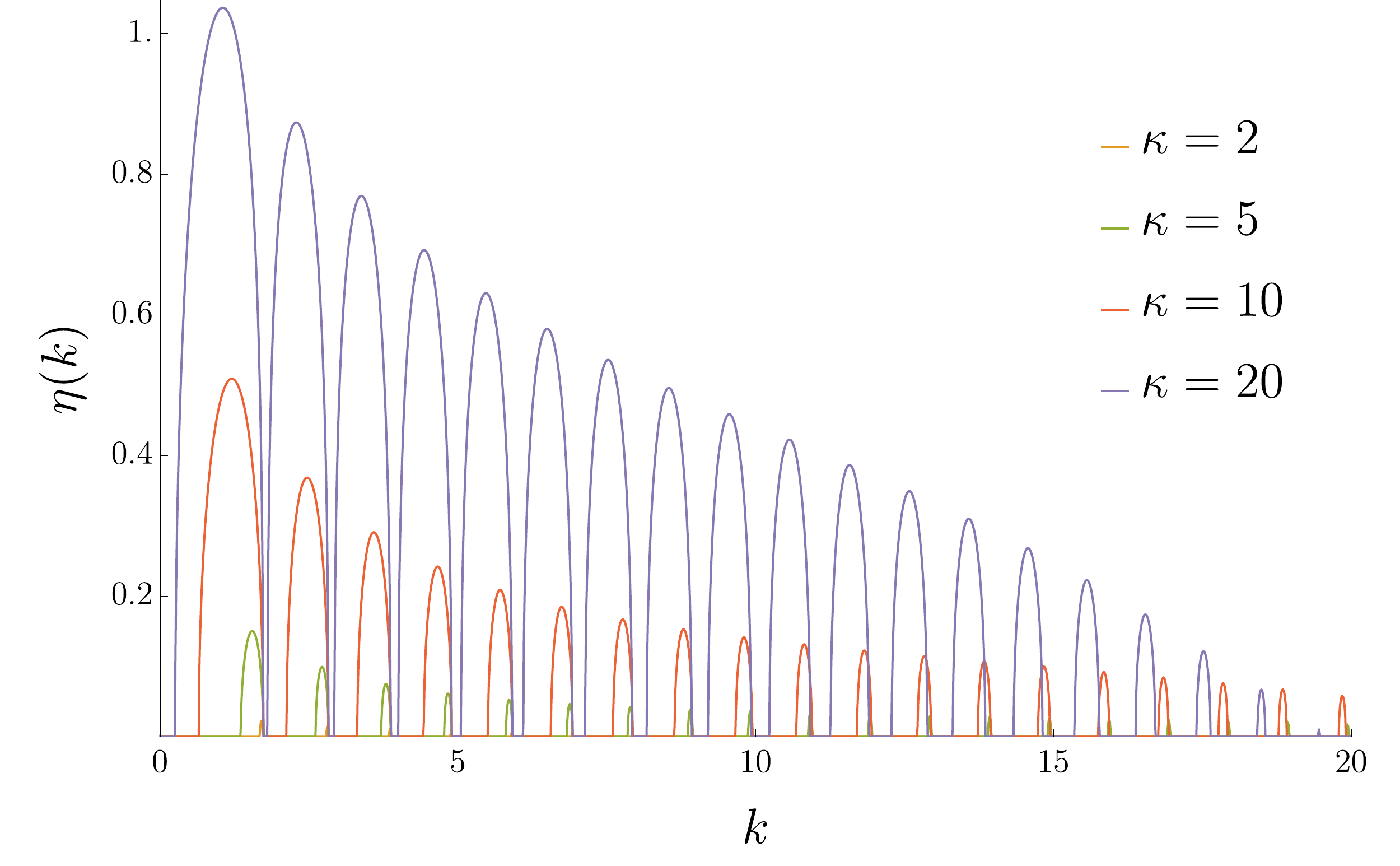}
   \label{fig:mrangemucombk20}
 }
\subfigure[
      ]
    {   \includegraphics[width=0.48\textwidth]{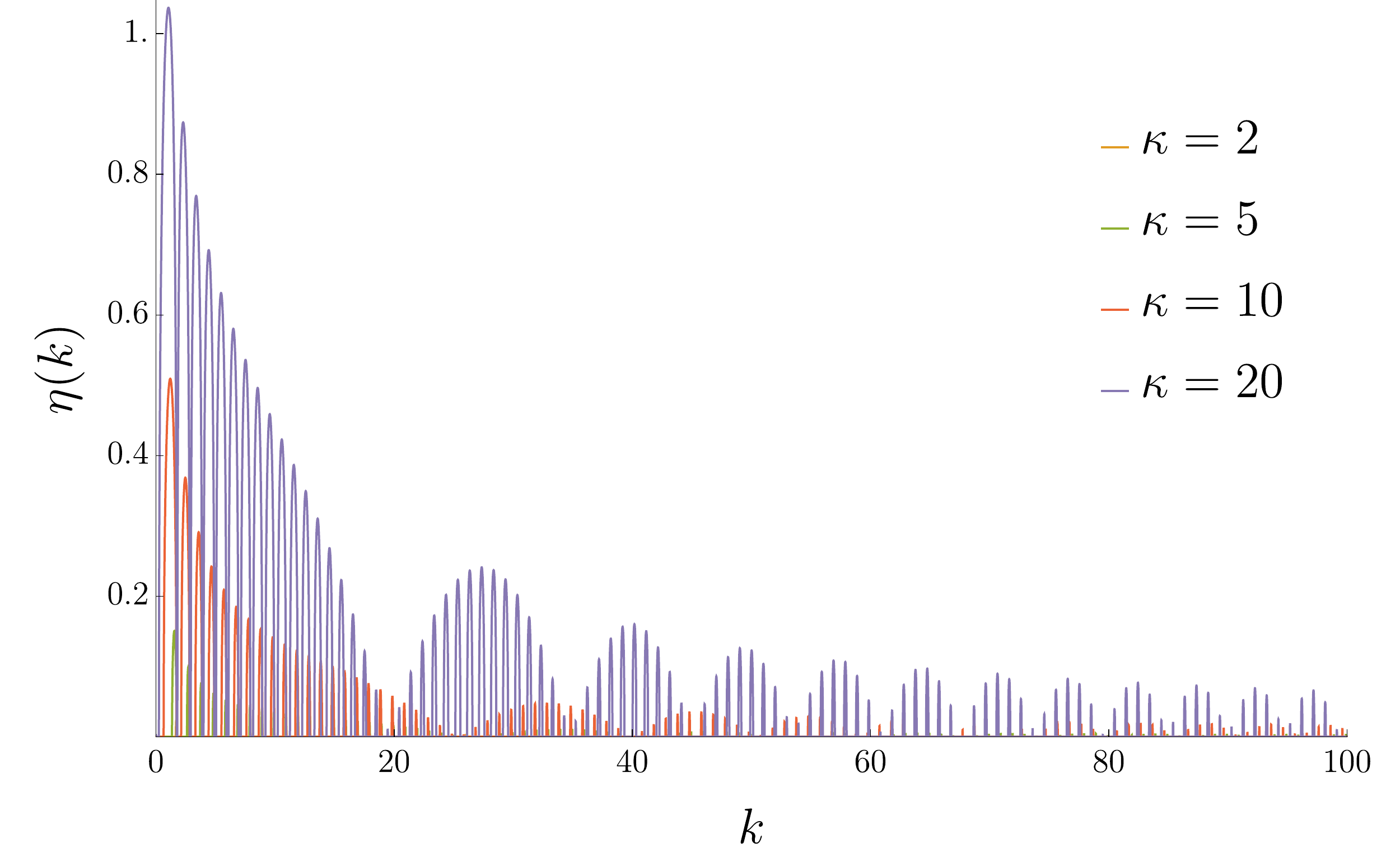}
   \label{fig:mrangemucombk100}
 }
 \centering\caption{Combination of the plots $\eta(k)$ vs.~$k$ for $\tfrac{\varphi_{\mathrm{initial}}}{2\pi}=100$ and  $\kappa=2.0, \,5.0, \,10, \, 20$.}
 \label{fig:mrangecombmuplots}
  \end{figure}

\begin{figure}
    \centering
      \subfigure[
    ]
    {   \includegraphics[width=0.48\textwidth]{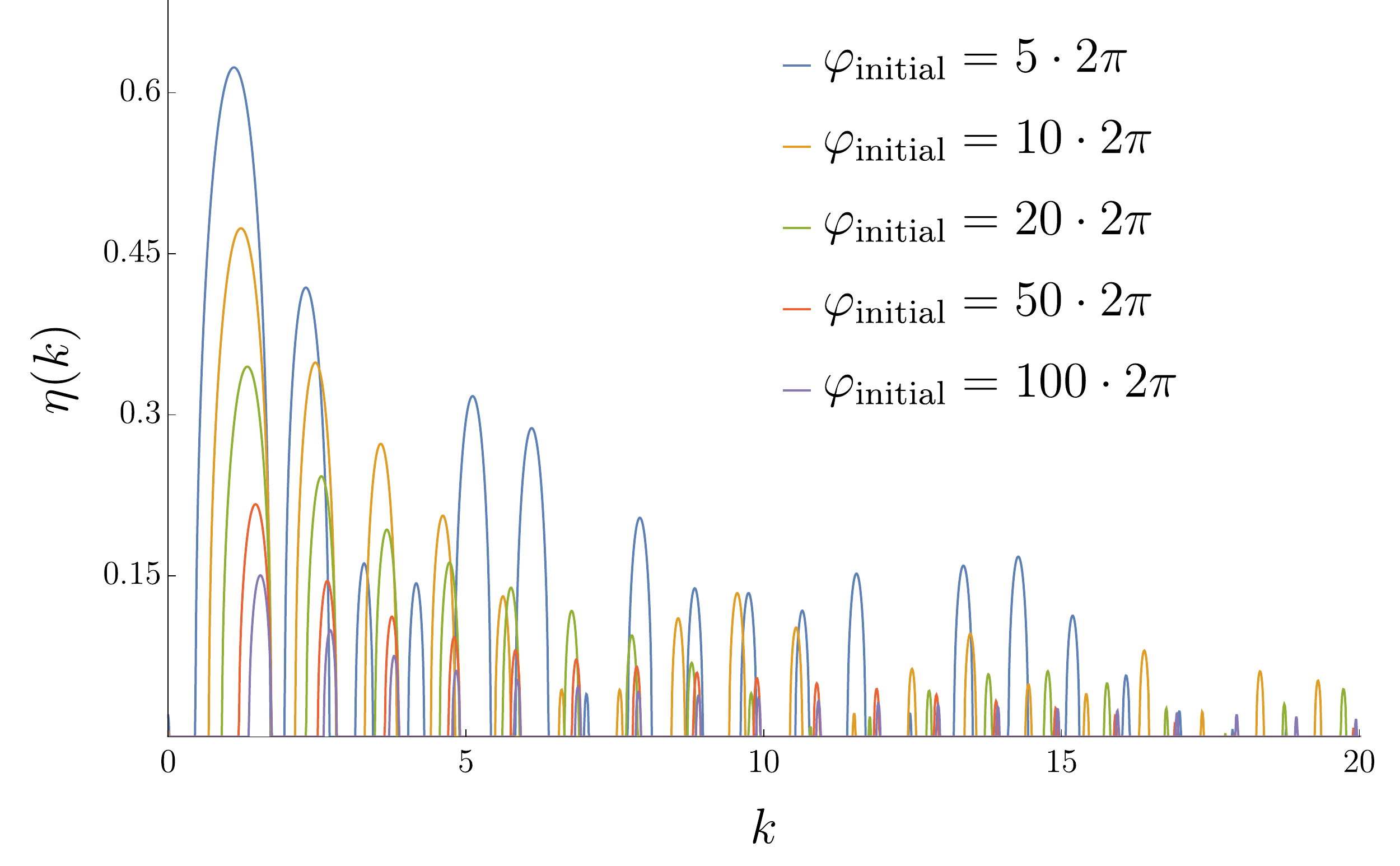}
   \label{fig:phirangemucombk20}
 }
\subfigure[
      ]
    {   \includegraphics[width=0.48\textwidth]{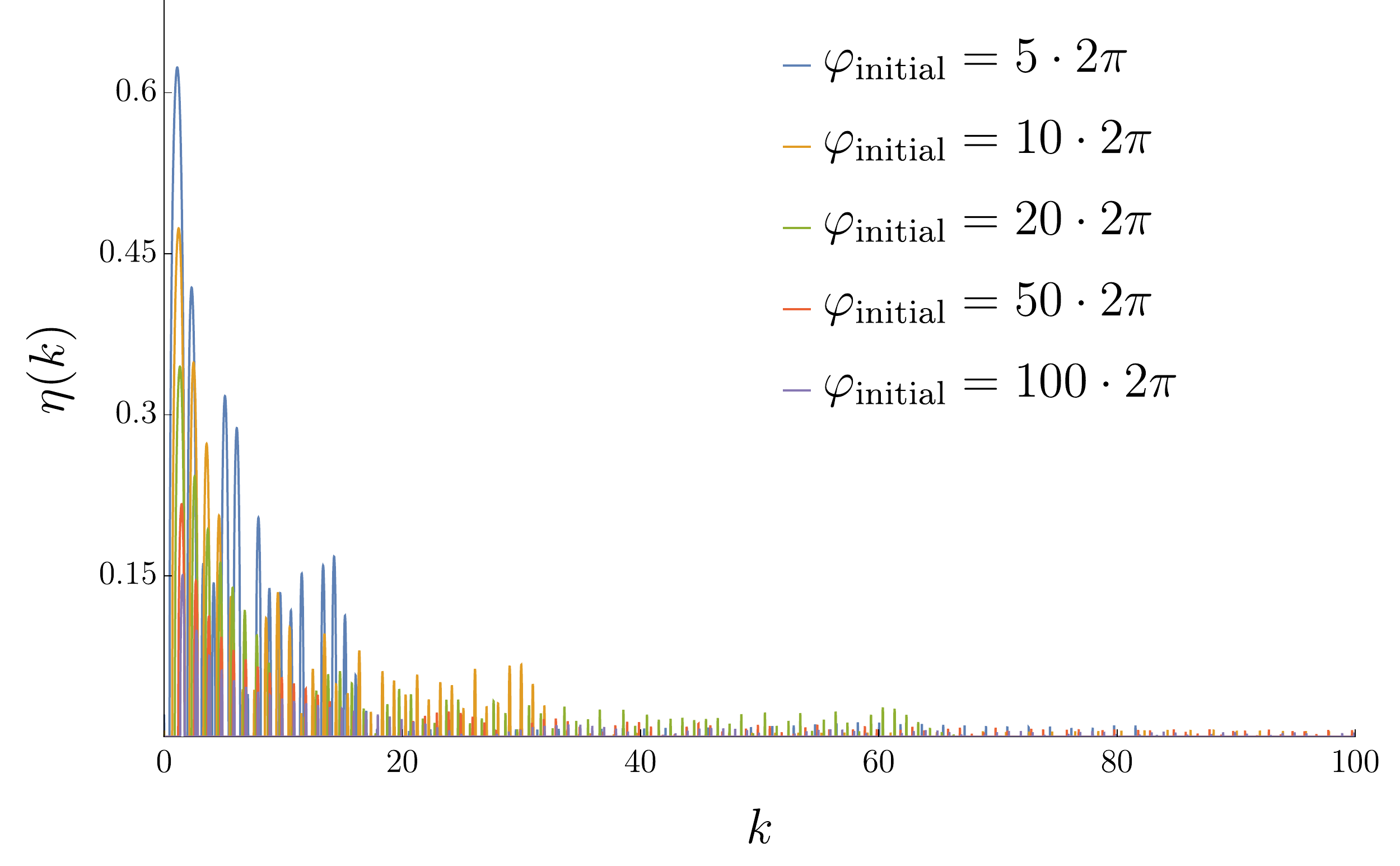}
   \label{fig:phirangemucombk100}
 }
 \centering\caption{Combination of plots of $\eta(k)$ vs.~$k$ for $\kappa=5.0$ and $\tfrac{\varphi_{\mathrm{initial}}}{2\pi}=5.0, \, 10, \, 20, \, 50, \, 100$.}
 \label{fig:phirangecombmuplots}
  \end{figure}

We obtain the following results. Depending on $k$ the solutions to \eqref{fluceom} exhibit alternating regions of stability or instability. Such a banded structure with regions of stability and instability is a well-known feature of solutions to Hill's equation (see e.g.~\cite{WhittakerWatson}). In regions of stability the solution is purely oscillatory and does not grow. In the instability bands the solution takes the form 
\begin{equation}
c_{\mathbf{k}}\sim \exp[\eta(k)\tau] \exp(-i\omega_{k} \tau) \ ,
\end{equation}
with $\eta(k) >0$ and  $\omega_k \sim n$ with $n \in \mathbbm{N}$. The growth exponent $\eta(k)$ is plotted in figs.~\ref{fig:mrangemuplots}, \ref{fig:phirangemuplots}, \ref{fig:mrangecombmuplots} and \ref{fig:phirangecombmuplots} for different values of $\kappa$ and $\varphi_{\mathrm{initial}}$.

As can be seen in figures \ref{fig:mrangemuplots} and \ref{fig:mrangecombmuplots} the width of instability regions as well as the size of the growth exponents increases when $\kappa$ is increased (for fixed $\varphi_{\mathrm{initial}}$). This can be understood as follows: increasing $\kappa$ enhances the tachyonic regions of the potential thus leading to more growth of fluctuations. This also implies that for sufficiently small $\kappa$ the growth of fluctuations is heavily suppressed or even absent. We find that fluctuations do not grow significantly for $\kappa \lesssim 1.0$.  

The growth of fluctuations is also affected by the value of $\varphi_{\mathrm{initial}}$. When $\varphi_{\mathrm{initial}}$ is lowered (for fixed $\kappa$) the growth of fluctuations is increased as can be seen in \ref{fig:phirangemuplots} and \ref{fig:phirangecombmuplots}. Again, both the width and height of instability bands is enhanced. This is also expected: for lower $\varphi_{\mathrm{initial}}$ the background field will spend a larger fraction of time per period in the tachyonic regions of the potential resulting in more growth of fluctuations.

Further, note that for the chosen parameters the growth of fluctuations can be very rapid compared to the oscillation period of the background field. Given that we measure time in units of $m_{\mathrm{mono}}$ the background field oscillates with a period of $\sim (2 \pi)$. We find that growth by an e-fold or more can occur within a few oscillations of the background solution. In that sense it is very rapid growth. 

Another observation is that higher modes are typically growing more slowly than the lower modes (cf.~figs.~\ref{fig:mrangemuplots} and \ref{fig:phirangemuplots}). Deviations from this behaviour can occur when the `wiggles' are very pronounced over the whole accessible field range, i.e.~for small $\varphi_{\mathrm{initial}}$.

The occurrence of growth of fluctuations can also be understood in terms of a resonance condition. Whenever growth occurs $c_{\mathbf{k}}$ oscillates with an angular frequency $\omega_{k} \in \mathbbm{N}$. As we have seen in the previous section~\eqref{classical} the background solution $\varphi_0(\tau)$ in this case oscillates with an angular frequency of $\sim 1$. With the above condition the fluctuation performs an integral number of oscillations for each oscillation of the background field. In this sense it is driven resonantly by the background field. 

Let us already make a comment on how the above findings change if we also allow the universe to expand. For one, both the amplitudes of the background solution and the fluctuations will be damped by the expansion. On the other hand, we also observed that fluctuations grow faster if the amplitude of oscillation of the background field is lowered. This effect will lead to an enhancement of fluctuations as the universe expands.

\subsection{Fluctuations in position space}
So far we have discussed the growth of fluctuations on a mode by mode basis, i.e.~in momentum space. 
In practice we are often more interested in the growth of the size of a fluctuation at a given point. 
The fluctuating field is given by the Fourier transform of all the $\mathbf{k}$-modes,
\begin{equation}
\varphi(\varix,\tau) = \varphi_{0}(\tau)+\int \frac{d^{3}k}{(2\pi)^3} \, c_{\mathbf{k}}(\tau) \, \exp(i\mathbf{k}\cdot\varix) \ .
\end{equation}
Taking only the leading order growth into account we have,
\begin{equation}
\varphi(\tau) \sim \varphi_{0}(\tau)+\int \frac{d^{3}k}{(2\pi)^3} \, c_{\mathbf{k}}(0) \, \exp(\eta(k)\tau) \, \exp(-i\omega_{\mathbf{k}}\tau) \, \exp(i\mathbf{k}\cdot\varix) \ .
\end{equation}

\begin{figure}
    \centering
      \subfigure[
    ]
    {   \includegraphics[width=0.44\textwidth]{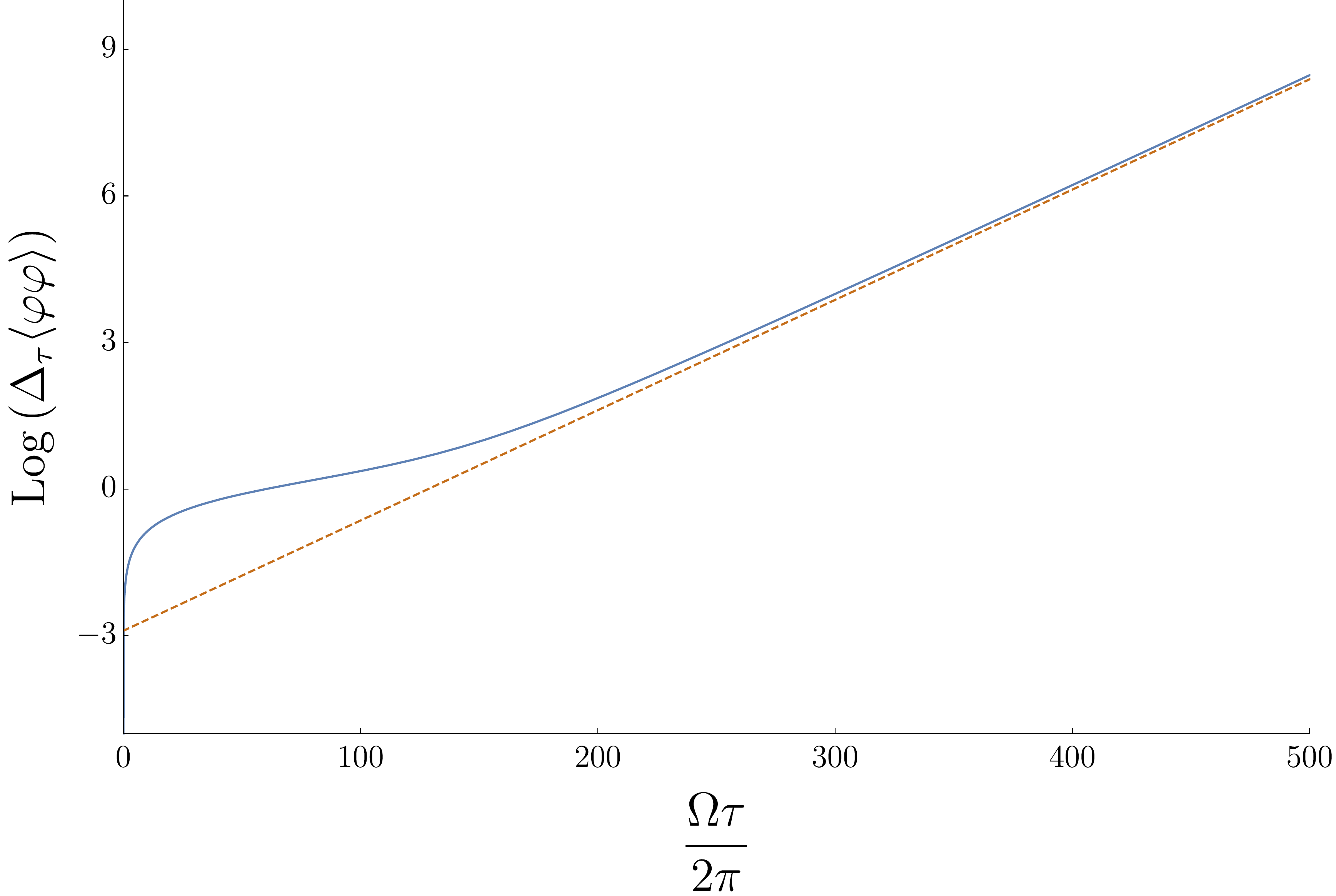}
   \label{fig:GrowthIntM1}
 }
  \subfigure[
    ]
    {   \includegraphics[width=0.44\textwidth]{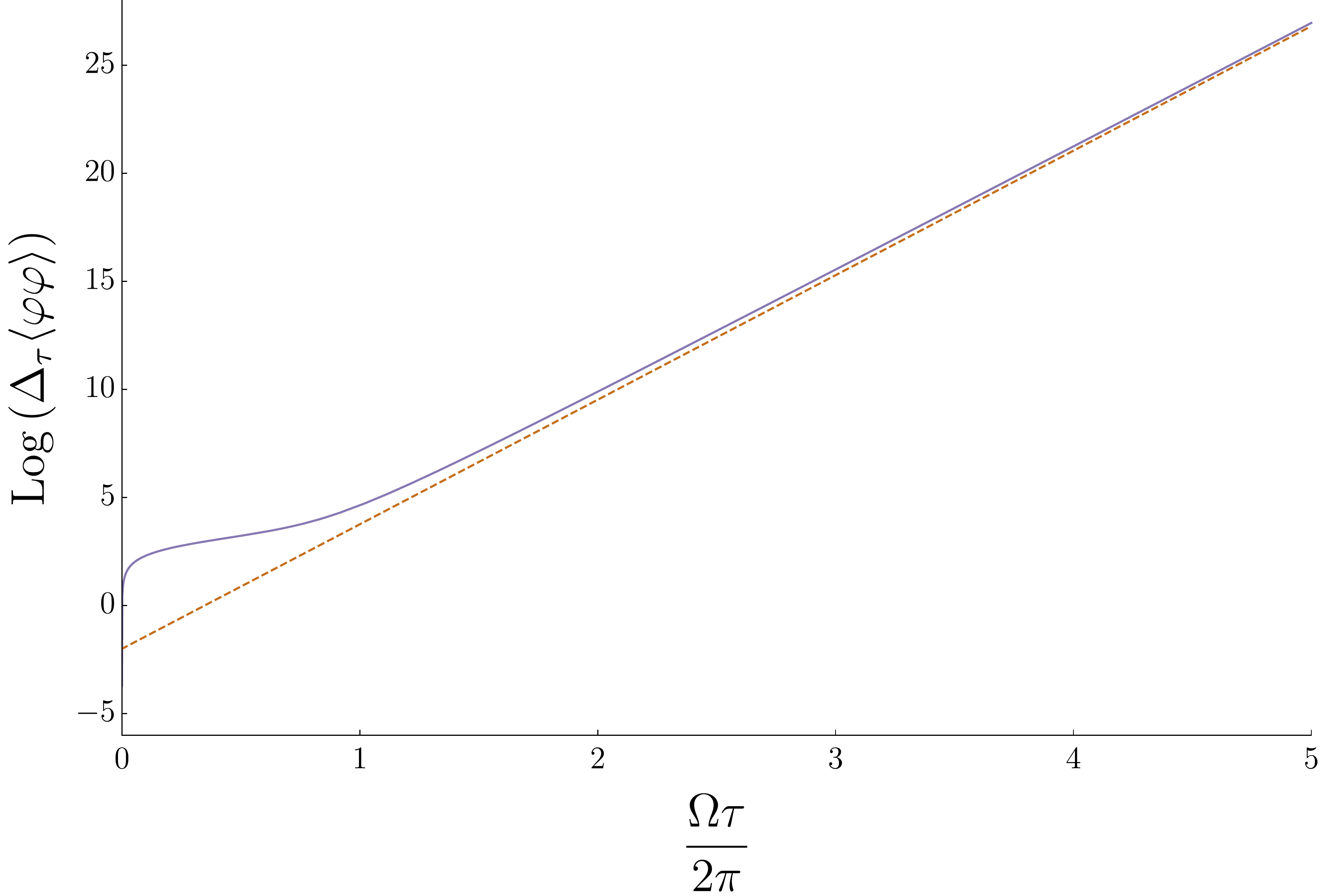}
   \label{fig:GrowthIntM20}
 }
 \\
      \subfigure[
    ]
    {   \includegraphics[width=0.44\textwidth]{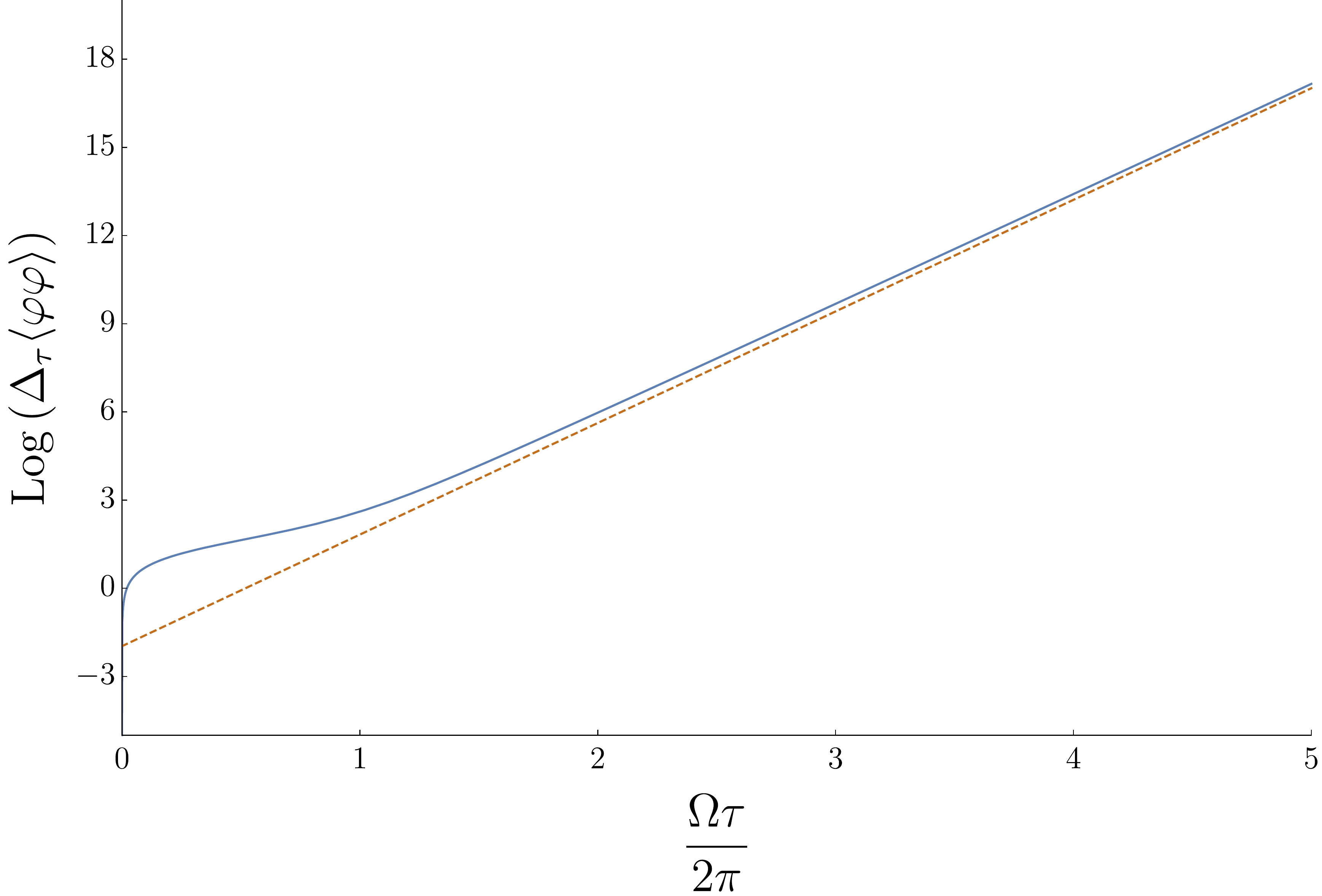}
   \label{fig:GrowthIntphii5}
 }
 \subfigure[
   ]
    {   \includegraphics[width=0.44\textwidth]{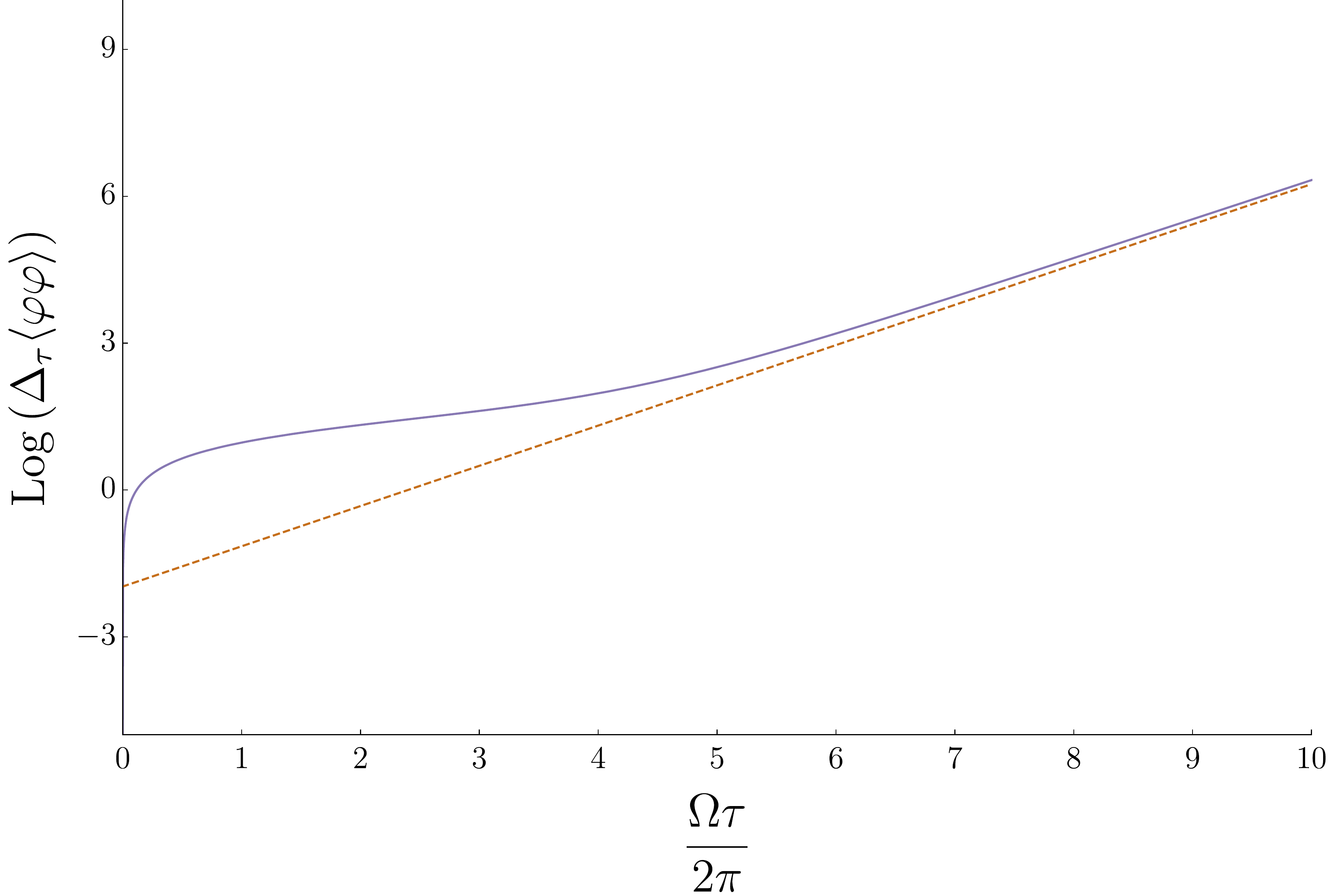}
   \label{fig:GrowthIntphii100}
 }
      \caption{Growth of the correlation function $\Delta_\tau \langle \varphi(\varix)\varphi(\varix)\rangle$ defined in \eqref{eq:growthfunc} vs.~the number of oscillations of the background solution $\varphi_0$ for (a) $\tfrac{\varphi_{\mathrm{initial}}}{2 \pi} =100$ and $\kappa=1.0$, (b) $\tfrac{\varphi_{\mathrm{initial}}}{2 \pi} =100$ and $\kappa=20$, (c) $\tfrac{\varphi_{\mathrm{initial}}}{2 \pi} =5$ and $\kappa=5.0$, (d) $\tfrac{\varphi_{\mathrm{initial}}}{2 \pi} =100$ and $\kappa=5.0$. The dashed lines are proportional to $\sim \exp (2 \eta_{\rm max} \tau)$. }
    \label{fig:integrated}
  \end{figure}

In many cases we can approximate the fluctuating modes by a Gaussian ensemble with a spectrum given by
\begin{equation}
\langle c_{\mathbf{k}}c^{\star}_{\mathbf{k}'}\rangle=P(k) (2\pi)^3\delta(\mathbf{k}-\mathbf{k}').
\end{equation}
To study the size of fluctuations at a point, an appropriate quantity is the correlation function. For example we can consider 
\begin{equation}
\langle \varphi(\varix)\varphi(\varix)\rangle(\tau) \sim \int \frac{d^{3}k}{(2\pi)^3}P(k,\tau=0) \exp(2\eta(k)\tau).
\end{equation}
For some initial spectra, like the quantum initial conditions we discuss below, one finds
\begin{equation}
\int \frac{d^{3}k}{(2\pi)^3}P(k,\tau=0)=\infty.
\end{equation}
Hence, strictly speaking, a renormalization condition has to be employed.
However, and more importantly, we have checked (numerically) that the {\emph{growth}} in the 
correlation function,
\begin{align}
\nonumber \Delta_{\tau} \langle \varphi (\varix)\varphi(\varix)\rangle & \equiv \langle \varphi (\varix)\varphi(\varix)\rangle(\tau)-\langle \varphi (\varix)\varphi(\varix)\rangle(0) \\
\label{eq:growthfunc} &\sim \int \frac{d^{3}k}{(2\pi)^3}P(k,\tau=0)\left[ \exp(2\eta(k)\tau)-1\right]<\infty
\end{align}
even for initial conditions (like the quantum ones),
\begin{equation}
P(k,\tau=0)\to {\rm const}\quad{\rm for}\quad k \to \infty.
\end{equation}
At late times the expression \eqref{eq:growthfunc} takes a very simple form. It is dominated by the mode $k_*$ corresponding to maximum value for the growth parameter $\eta_{\rm max} \equiv \eta(k_*)$:
\begin{equation}
\Delta_{\tau} \langle \varphi(\varix)\varphi(\varix)\rangle \sim \exp (2 \eta_{\rm max} \tau) \ .
\end{equation}
In Fig.~\ref{fig:integrated} we display the growth in the correlation function \eqref{eq:growthfunc} for constant $P(k,\tau=0)=1$.

\subsection{Initial quantum fluctuations and the impact on cosmology}\label{quantumini}
In principle initial conditions at a given time will depend on the previous evolution of the Universe.
A period of inflation asks for different initial conditions than a static situation. 

To get some idea of a typical, minimal amount of fluctuations let us consider the fluctuations inherent in a 
relativistic quantum field fluctuating around its true vacuum state, where we also neglect interaction effects.
This can be described as a set of independent harmonic oscillators for each $\mathbf{k}$-mode (we closely follow~\cite{1503.02907} where one can also find more details). At the initial time $t_{0}$ we have,
\begin{equation}
\phi(\mathbf{x},t_{0})=\phi(t_{0})+\int \frac{d^{3}p}{(2\pi)^3} \sqrt{\frac{f_{\mathbf{p}}+1/2}{\omega_{\mathbf{p}}}}
\, a_{\mathbf{p}} \, \exp(i\mathbf{p}\cdot\mathbf{x}),
\end{equation}
where
\begin{equation}
\omega_{\mathbf{p}}=\sqrt{m^2+|\mathbf{p}|^2}.
\end{equation}
The $f_{\mathbf{p}}$ are the occupation numbers of each mode and the $1/2$ give the quantum fluctuations 
of each mode.
A similar equation holds for the time derivative,\footnote{As already mentioned in our evaluation of the fluctuation growth, we ignore the initial condition for the time derivative of the fluctuation modes. This changes the absolute values by ${\mathcal{O}}(1)$ factors which we neglect in light of the enormous exponential growth.}  $\dot{\phi}$.

The coefficients $a_{\mathbf{p}}$ are given by Gaussian random numbers  with random phases normalized such that,
\begin{equation}
\langle a^{\star}_{\mathbf{p}}a_{\mathbf{p'}}\rangle =(2\pi)^3\delta(\mathbf{p}-\mathbf{p}'),\quad
\langle a_{\mathbf{p}}a_{\mathbf{p'}}\rangle =0,\quad \langle a^{\star}_{\mathbf{p}}a^{\star}_{\mathbf{p'}}\rangle =0.
\end{equation}
For our case of a real field one also requires,
\begin{equation}
a^{\star}_{\mathbf{k}}=a_{-\mathbf{k}}.
\end{equation}
Performing the appropriate re-scalings to the dimensionless variables~\eqref{eq:rescale} we have for the correlation
function,
\begin{equation}
\langle\varphi(\varix)\varphi(\varix)\rangle=\frac{m^{2}_{\mathrm{mono}}}{f^2}\int \frac{d^{3} k}{(2\pi)^3} \frac{f_{k}+1/2}{\omega_{k}}.
\end{equation}
Comparing with our definition of $P(k)$ we therefore have,
\begin{equation}
P(k)=\frac{m^{2}_{\mathrm{mono}}}{f^2}\left(f_{\mathbf{k}}+\frac{1}{2}\right).
\end{equation}
For the case of an initially non-excited quantum state we therefore have,
\begin{equation}
\label{sizefluc}
P(k)=\frac{m^{2}_{\mathrm{mono}}}{2f^2},
\end{equation}
independent of $\mathbf{k}$.

While the use of the classical equations of motion is not fully appropriate when the occupation numbers are small
the discussion of the previous sections tells us that fluctuations and therefore occupations numbers grow rapidly.
Therefore we soon enter a regime where the use of the classical field equations is justified and should describe the evolution well.

Eq.~\eqref{sizefluc} tells us that for large $f$ and small values of $m_{\mathrm{mono}}$ the initial fluctuations are indeed 
very small, as one might expect.
Inserting typical values of interest for dark matter we have
\begin{equation}
P(k)\sim 10^{-50}\,\left(\frac{10^{10}\,{\rm GeV}}{f}\right)^{2}\left(\frac{m_{\mathrm{mono}}}{\mu{\rm eV}}\right)^{2}.
\end{equation}
This is indeed very small. Nevertheless, for the growth observed in the numerical examples above we only
need a few hundred oscillation periods for fluctuations to be of order $1$ (in our units of the periodicity scale $f$).

This raises the question of how many oscillations we can have in a realistic model. This of course depends on the expansion of the Universe.
Roughly speaking the inverse of the dimensionless Hubble parameter gives the number of oscillation times
before the Universe has expanded by one e-fold. 

One consequence of  expansion is the dilution of the fields. Thus the amplitude of oscillation of the background solution will decrease with the volume expansion. Similarly, the amplitude of fluctuations will also be lowered. However, there is a further effect on the fluctuations. Recall that the growth of fluctuations increases if the background field oscillates with a smaller amplitude. Hence, by lowering the amplitude of the background solution the expansion of the universe will eventually lead to an increased growth of fluctuations.

Another important effect for our purposes is that the physical $\mathbf{k}$ is rescaled by the expansion parameter,
\begin{equation}
{\mathbf{k}}\rightarrow {\mathbf{k}}/a.
\end{equation}
Accordingly, each mode effectively moves from high $|\mathbf{k}|$ to low $|\mathbf{k}|$ and only
spends a time,
\begin{equation}
\delta\tau \sim \log\left(1+\frac{\Delta k_{\rm res}}{k_{\rm res}}\right)\frac{1}{h}
\end{equation}
in each instability band centred on $k_{\rm res}$ with width $\Delta k_{\rm res}$. 

When the amplitude of the background field is such that the growth is strong, the width is also relatively wide (cf.~Fig.~\ref{fig:m20muk20}) such that
\begin{equation}
\frac{\Delta k_{\rm res}}{k_{\rm res}}\sim 1.
\end{equation}
Therefore if such a regime is entered when
\begin{equation}
h\ll 10^{-2}
\end{equation}
fluctuations will become important and cannot be neglected anymore. At this point, of course our approximation breaks down and a full field theory simulation is required. 

\section{Conclusions}\label{conclusions}
In this paper we propose fields exhibiting a so-called (axion) monodromy as viable Dark Matter candidates.
Compared to pNGBs the field range as well as the range of potential values is significantly
enlarged. This opens up wide ranges of parameter space for dark matter searches. In particular, for small dark matter masses the coupling to ordinary matter can be significantly enhanced compared to pseudo-Nambu-Goldstone dark matter candidates. For the important example of dark matter from axion-like particles coupled to two photons, the existence of a monodromy opens up previously inaccessible regions of parameter space (see Fig.~\ref{parameterspace}). This provides motivation for experimental searches in regions that are in principle accessible by experiment, but up to now lacked theoretical motivation for experimental scrutiny.

Beyond this result the potential of monodromic scalar fields can exhibit much more structure then a pure cosine or quadratic potential
of a pNGB (cf.~Fig.~\ref{potsplot}). This leads to a rich phenomenology (but also potential pitfalls where the model can be unsuitable for cosmology) of which we only have scratched the surface. In the following we list some of the features that we encountered.
\begin{itemize}
\item{}Depending on the initial values and the following cosmological evolution, the dark matter field may settle in different local minima,
corresponding to different values of the cosmological constant and also different masses of the dark matter particle.
\item{}When the field spends considerable time in the ``wiggly'' part of the potential (cf. Fig.~\ref{potsplot})
the equation of state can deviate significantly from the dark matter value $w=0$. 
\end{itemize}
The two observations above concern the classical field evolution. However, there are further important effects.
\begin{itemize}
\item{} We find that fluctuations of a characteristic size determined by the mass scale of the scalar field $L\sim 2\pi/m_{\mathrm{mono}}$
can grow exponentially and can become important.
\end{itemize}
All these features are important for the dark matter phenomenology. Interestingly, the effects described above are a direct consequence of the structure of the monodromy potential (polynomial + wiggles) and the rate of expansion, but are not limited to applications to dark matter phenomenology. Therefore, similar effects may also arise in other situations where monodromy potentials appear in cosmology, such as inflation.

Last, let us sketch a few directions for further study.
For applications to dark matter there is a clear need for a more elaborate study of the viable parameter space. In particular, a more detailed analysis of the allowed parameter ranges that are permitted by the changes of the equation of state would be desirable.
Further, it would also be interesting to determine when fluctuations become important to not only affect the equation of state but also structure formation.
This latter question is also intrinsically linked to obtaining a suitable description in the regime of large fluctuations in which 
our linear approximation breaks down. One possible path towards this goal may be numerical solutions of the classical field equations. Initial studies in the direction are under way~\cite{Redondoprivate}.

\section*{Acknowledgements}
The authors would like to thank A. Hebecker, E.~Kiritsis, J.~Redondo, F. Rompineve and A. Westphal for interesting discussions on monodromies and their phenomenology. We also gratefully acknowledge support from the DFG transregional research collaborative TR33 ``The Dark Universe''.

\appendix
\section{Hill's determinental method}
\label{sec:Hill}
In the following we will describe Hill's determinental method applied to the equation of motion for the fluctuations \eqref{fluceom}. Using this method we can obtain an expression for the instability parameter governing the growth of the solutions to \eqref{fluceom}.

The first observation is that the equation of motion for the fluctuation \eqref{fluceom} is a Hill's equation, i.e.~it is a second order differential equation with periodic coefficients. In our case it is given by \eqref{fluceom}:
\begin{align}
\label{eq:fluceomapp} & \ddot{c}_{\mathbf{k}} + p_k(\tau) c_{\mathbf{k}}=0 \ , \qquad \textrm{with} \\
\nonumber & p_k(\tau) = 1+ k^2 + \kappa^2 \cos \varphi_0(\tau) \ ,
\end{align}
The background solution has period $2 \pi / \Omega$ (see section \ref{classical}). It follows that $p_k(\tau)$ is periodic with period $\pi / \Omega$. In the following it will be convenient to introduce the variable $z = \Omega \tau$. Then $p_k$ has period $\pi$ as a function of $z$. In the remainder of this section we will closely follow the analysis in \cite{WhittakerWatson} for Hill's equation.

First, we express $\cos (\varphi_0(z/ \Omega))$ as a Fourier series:
\begin{equation}
\cos (\varphi_0(z / \Omega)) = \frac{a_0}{2} + \sum_{n=1}^\infty a_n \cos (2 n z) \ .
\end{equation}
Interestingly, once we insert a solution for $\varphi_0$ the series truncates at some $n=N$. We can then write the equation of motion for the fluctuation as
\begin{align}
\label{eq:Hill}
& {c}_{\mathbf{k}}^{\prime \prime} + \left[\theta_0 + \sum_{n=1}^N 2 \theta_n \cos (2 n z) \right] {c}_{\mathbf{k}}=0 \ , \\
\nonumber & \theta_0= \Omega^{-2} \left(1 + k^2 + \kappa^2 \frac{a_0}{2} \right) \ , \\
\nonumber & \theta_n= \Omega^{-2} \kappa^2 \frac{a_n}{2} \ , \qquad n \neq 0 \ ,
\end{align}
where `prime' denotes a derivative w.r.t.~$z$. 

According to Floquet theory this equation will have solutions of the form $e^{\mu z} F(z)$, where $F(z)$ should be a periodic function with the same period as $p_k(z)$, i.e.~$F(z + \pi) = F(z)$. The next step is to write this candidate solution as a Laurent expansion
\begin{equation}
e^{\mu z} F(z) = e^{\mu \tau} \sum_{m=- \infty}^{\infty} b_m e^{2 i m z}
\end{equation}
Inserting this expansion into \eqref{eq:Hill} then gives rise to a recursion relation for the coefficients $b_{m}$:
\begin{equation}
(i \mu- 2 m)^2 b_{m} - \sum_{n=-N}^N \theta_n b_{m-n} =0 \ ,
\end{equation}
where we defined $\theta_{-n} \equiv \theta_n$. For reasons of convergence it will be useful to divide this by $(4 m^2-\theta_0)$. Then the recursion relation can also be written as a matrix equation $D_{m n} b_n =0$ with
\begin{equation}
D_{mn} (i \mu)= \frac{(i \mu- 2 m)^2}{4m^2 -\theta_0} \delta_{mn} - \frac{\theta_{m-n}}{4m^2 - \theta_0} \ ,
\end{equation}
where $m, n = -\infty , \ldots , \infty$.
The determinant $\Delta(i \mu) \equiv |D_{mn} (i \mu)|$ is called Hill's determinant and it should vanish for a solution. The equation $\Delta(i \mu)=0$ is then an implicit equation for $\mu$. As detailed in \cite{WhittakerWatson} one can show that 
\begin{equation}
\Delta(i \mu)= \Delta(0) - \frac{\sin^2 (\tfrac{\pi}{2} i \mu)}{\sin^2 (\tfrac{\pi}{2} \sqrt{\theta_0})} \ .
\end{equation}
Thus, the solutions to Hill's determinental equation $\Delta(i \mu)=0$ are given by
\begin{equation}
\label{eq:musol}
\sin^2 (\tfrac{\pi}{2} i \mu) = \Delta(0) \sin^2 (\tfrac{\pi}{2} \sqrt{\theta_0}) \ ,
\end{equation}
with
\begin{equation}
\Delta(0) = \begin{vmatrix}
  \ddots & \vdots & \vdots & \vdots & \vdots & \vdots & \reflectbox{$\ddots$} \\
  \cdots & 1 & \frac{- \theta_1}{16 - \theta_0} & \frac{- \theta_2}{16 - \theta_0} & \frac{- \theta_3}{16 - \theta_0} & \frac{- \theta_4}{16 - \theta_0}  & \cdots \\
  \cdots & \frac{- \theta_1}{4 - \theta_0} & 1 & \frac{- \theta_1}{4 - \theta_0} & \frac{- \theta_2}{4 - \theta_0} & \frac{- \theta_3}{4 - \theta_0} & \cdots \\
  \cdots & \frac{- \theta_2}{0 - \theta_0} &\frac{- \theta_1}{0 - \theta_0} & 1 & \frac{- \theta_1}{0 - \theta_0} & \frac{- \theta_2}{0 - \theta_0}  & \cdots \\
  \cdots & \frac{- \theta_3}{4 - \theta_0} &  \frac{- \theta_2}{4 - \theta_0} & \frac{- \theta_1}{4 - \theta_0} & 1 & \frac{- \theta_1}{4 - \theta_0} & \cdots \\
  \cdots & \frac{- \theta_4}{16 - \theta_0} & \frac{- \theta_3}{16 - \theta_0} & \frac{- \theta_2}{16 - \theta_0} & \frac{- \theta_1}{16 - \theta_0} & 1 & \cdots \\
 \reflectbox{$\ddots$} & \vdots & \vdots & \vdots & \vdots & \vdots & \ddots \\
 \end{vmatrix} \ .
\end{equation}

Using eq.~\eqref{eq:musol} we can calculate the growth parameter $\mu$ for different values of $k=|\mathbf{k}|$. To this end we need to evaluate the determinant $\Delta(0)$ of the infinite matrix $D_{mn}(0)$. This matrix has the following properties. As the Fourier coefficients $a_n$ of $\cos \varphi_0$ vanish for some $n>N$ it follows that $\theta_n=0$ for $n > N$. Thus the matrix $D_{mn}(0)$ only has non-zero components in a band of width $2N+1$ centred on the diagonal. Furthermore, the off-diagonal entries become smaller if the move away from the central row.  An approximate expression for $\Delta(0)$ can then be found by only considering the determinant of the finite square matrix centred on the element $D_{00}$. One can obtain an increasingly good approximation by increasing the size of this finite matrix.

As we are interested in the growth of fluctuations in time $\tau$, let us define $\eta \equiv \Omega \mu$. The result of the analysis in this section is that fluctuations $c_{\mathbf{k}}$ evolve as 
\begin{equation}
c_{\mathbf{k}} \sim e^{\mu (k) z} F(z) = e^{\eta(k) \tau} F(\Omega \tau) \ , \qquad F(\Omega \tau+ \pi) = F(\Omega \tau) \ ,
\end{equation}
where $\eta(k)$ is calculated as described above. The results for $\eta(k)$ shown in figures \ref{fig:mrangemuplots}, \ref{fig:phirangemuplots}, \ref{fig:mrangecombmuplots} and \ref{fig:phirangecombmuplots} were obtained using this method.

Let us close with a few further observations regarding the solutions to \eqref{eq:fluceomapp}. As a function of $k=|\mathbf{k}|$ the solution for $c_{\mathbf{k}}$ exhibits regions of stability or instability (see figures \ref{fig:mrangemuplots} and \ref{fig:phirangemuplots}). More specifically, for some ranges of $k$ the parameter $\eta(k)$ has a vanishing real part $\textrm{Re} (\eta(k))=0$ and fluctuations do not grow. In other parameter ranges of $k$ we find $\textrm{Re} (\eta(k))>0$ and fluctuations grow. This is typical for solutions to Hill's equation.

In addition, we compared the instability coefficient $\eta(k)$ obtained using Hill's determinant with those obtained from a direct numerical solution of \eqref{eq:fluceomapp} and find very good agreement. Using this numerical solution one can also make the following observation regarding the frequency of oscillation of the fluctuations. The maxima of the peaks in $\eta(k)$ occur when $\theta_0 \equiv \Omega^{-2} (1+k^2 + \kappa^2 \tfrac{a_0}{2})= n^2$ with $n \in \mathbbm{N}$. This can be explained as follows. When this condition is met the fluctuation $c_{\mathbf{k}}$ oscillates as $e^{i \Omega \sqrt{\theta_0} \tau}$ and is resonantly driven by the higher Fourier modes of $\cos(\varphi_0(\tau))$. Again, this behaviour is typical for solutions to Hill's equation.\footnote{Note that the instability band with $\theta_0 = 1$ can be absent when $a_0>0$. Then $\theta_0= \Omega^{-2}(1+ k^2 + \kappa^2 \frac{a_0}{2}) > 1$ and there is no $k$ such that $\theta_0 = 1$.}

\end{document}